\documentclass[sigplan]{acmart}
\renewcommand\footnotetextcopyrightpermission[1]{}
\usepackage{booktabs}
\usepackage{amsmath}
\usepackage{empheq}
\usepackage{algorithm}
\usepackage[vlined,linesnumbered,ruled,algo2e]{algorithm2e}
\usepackage{color}
\usepackage[labelfont=bf]{caption}
\usepackage{booktabs}
\usepackage{url}
\usepackage[normalem]{ulem}
\usepackage{verbatim}
\usepackage{listings}
\usepackage{adjustbox}
\usepackage{hyperref}
\usepackage{multirow}
\usepackage{makecell}
\usepackage{tabularx}
\usepackage[flushleft]{threeparttable}
\usepackage{subcaption}
\usepackage{soul}
\usepackage{fancyhdr}

\usepackage{pifont}

\usepackage{titlesec}
\titlespacing*{\section}{0pt}{0.5\baselineskip}{0.5\baselineskip}
\titlespacing*{\subsection}{0pt}{0.5\baselineskip}{0\baselineskip}

\definecolor{codegreen}{rgb}{0.0,0.5,0.0}
\definecolor{codegray}{rgb}{0.5,0.5,0.5}
\definecolor{codepurple}{rgb}{0.58,0,0.82}
\definecolor{codeblue}{rgb}{0.0, 0.5, 1.0}
\definecolor{skyblue}{rgb}{0.67, 0.9, 0.93}
\lstdefinestyle{mystyle}{
    basicstyle={\linespread{0.9}\scriptsize\ttfamily},
    commentstyle=\color{codegreen},
    keywordstyle=\color{black},
    numberstyle=\tiny\color{codegray},
    breakatwhitespace=false,
    captionpos=b,
    keepspaces=true,
    numbers=left,
    numbersep=5pt,
    showspaces=false,
    showstringspaces=false,
    showtabs=false,
    tabsize=2,
    columns={fullflexible},
    gobble={4}
}
\newenvironment{tightlist}[4][\textbullet]
{
  \begin{list}{#1}
  {
    \setlength{\leftmargin}{#2}
    \setlength{\rightmargin}{#3}
    \setlength{\topsep}{0pt}
    \setlength{\parsep}{0pt}
    \setlength{\listparindent}{0pt}
    \setlength{\itemsep}{#4}
  }
}{
  \end{list}
}

\newcommand{\showcomments}{yes}

\newcommand{\fixme}[1]{
    \ifthenelse{\equal{\showcomments}{yes}}{\textcolor{red}{[fixme: #1]}}{\ignorespaces}
}

\newcommand{\zz}[1]{
    \ifthenelse{\equal{\showcomments}{yes}}{\textcolor{blue}{[zz: #1]}}{\ignorespaces}
}

\newcommand{\jj}[1]{
    \ifthenelse{\equal{\showcomments}{yes}}{\textcolor{cyan}{[jj: #1]}}{\ignorespaces}
}

\newcommand{\yh}[1]{
    \ifthenelse{\equal{\showcomments}{yes}}{\textcolor{purple}{[yh: #1]}}{\ignorespaces}
}

\begin{document}

\title{Analysis and Optimization of GNN-Based Recommender Systems on Persistent Memory}

\settopmatter{authorsperrow=4}

\author{Yuwei Hu}
\affiliation{
  \institution{Cornell University}
  \city{Ithaca, New York}
  \country{USA}
}
\email{yh457@cornell.edu}

\author{Jiajie Li}
\affiliation{
  \institution{Cornell University}
  \city{Ithaca, New York}
  \country{USA}
}
\email{jl4257@cornell.edu}

\author{Zhongming Yu}
\affiliation{
  \institution{UCSD}
  \city{San Diego, California}
  \country{USA}
}
\email{zhy025@ucsd.edu}

\author{Zhiru Zhang}
\affiliation{
  \institution{Cornell University}
  \city{Ithaca, New York}
  \country{USA}
}
\email{zhiruz@cornell.edu}

\begin{abstract}
Graph neural networks (GNNs), which have emerged as an effective method for handling machine learning tasks on graphs, bring a new approach to building recommender systems, where the task of recommendation can be formulated as the link prediction problem on user-item bipartite graphs.
Training GNN-based recommender systems (GNNRecSys) on large graphs incurs a large memory footprint, easily exceeding the DRAM capacity on a typical server.
Existing solutions resort to distributed subgraph training, which is inefficient due to the high cost of dynamically constructing subgraphs and significant redundancy across subgraphs.

The emerging persistent memory technologies provide a significantly larger memory capacity than DRAMs at an affordable cost, making single-machine GNNRecSys training feasible, which eliminates the inefficiencies in distributed training.
One major concern of using persistent memory devices for GNNRecSys is their relatively low bandwidth compared with DRAMs.
This limitation can be particularly detrimental to achieving high performance for GNNRecSys workloads since their dominant compute kernels are sparse and memory access intensive.
To understand whether persistent memory is a good fit for GNNRecSys training, we perform an in-depth characterization of GNNRecSys workloads and a comprehensive analysis of their performance on a persistent memory device, namely, Intel Optane.
Based on the analysis, we provide guidance on how to configure Optane for GNNRecSys workloads.
Furthermore, we present techniques for large-batch training to fully realize the advantages of single-machine GNNRecSys training.
Our experiment results show that with the tuned batch size and optimal system configuration, Optane-based single-machine GNNRecSys training outperforms distributed training by a large margin, especially when handling deep GNN models.

\end{abstract}

\maketitle

\thispagestyle{fancy}
\fancyhead{}

\vspace{-1em}
\section{Introduction}
\label{sec:introduction}




Graph neural networks (GNNs) have emerged as a promising approach to building recommender systems, where the task of recommendation can be formulated as the link prediction problem on user-item bipartite graphs \cite{ngcf, gcmc, pinsage, gnn-recsys-survey}.
More concretely, a GNN model generates a dense vector representation for each vertex (namely, the embedding vector), and a pair of vertex embeddings are used to carry out the per-edge prediction \cite{word2vec, deepwalk, node2vec}.
GNNs can generate high-quality vertex embeddings by incorporating multi-hop neighborhood information through iterative message passing \cite{gnn-chemistry, graphnets, gcn}.
For these reasons, recent years have seen a rapid surge of development on GNN-based recommender systems (GNNRecSys) in both academia and industry \cite{gcmc, ngcf, lightgcn, graphrec, pinsage, aligraph, uber-gnn}.

One major challenge faced by GNNRecSys workloads is that they consume a large amount of memory.
First, real-world user-item bipartite graphs are large.
For example, the Pinterest graph \cite{pinsage} has three billion vertices and eighteen billion edges.
Second and more importantly, the memory footprint of GNNs is incurred not only by the graph structure but also by embeddings on vertices and messages on edges.
As a result, GNNs consume 2--3 orders of magnitude more memory than traditional graph processing workloads such as PageRank, where each vertex/edge is associated with a scalar.
Even on a medium-size graph with three hundred million edges, training a three-layer GNN model with the embedding length set to 128 would require 500 GB of memory, easily exceeding the DRAM capacity on a typical server.

To tackle the memory capacity bottleneck, existing efforts resort to distributed GNN training \cite{distgnn, distdgl, p3}, among which DistDGL \cite{distdgl} is the only open-source framework.
DistDGL adopts a subgraph training approach --- on each machine, it selects a batch of target vertices, constructs a subgraph containing all the vertices/edges required to compute the embeddings for the target vertices, and then performs training on the subgraph.
This subgraph training approach, however, is inefficient because (1) the cost of dynamically constructing subgraphs for every batch is high, and (2) different subgraphs have overlaps, causing redundancy in both computation and memory consumption.

\begin{figure*}[t]
\centering

\begin{subfigure}[b]{0.2\linewidth}
\includegraphics[width=0.9\linewidth]{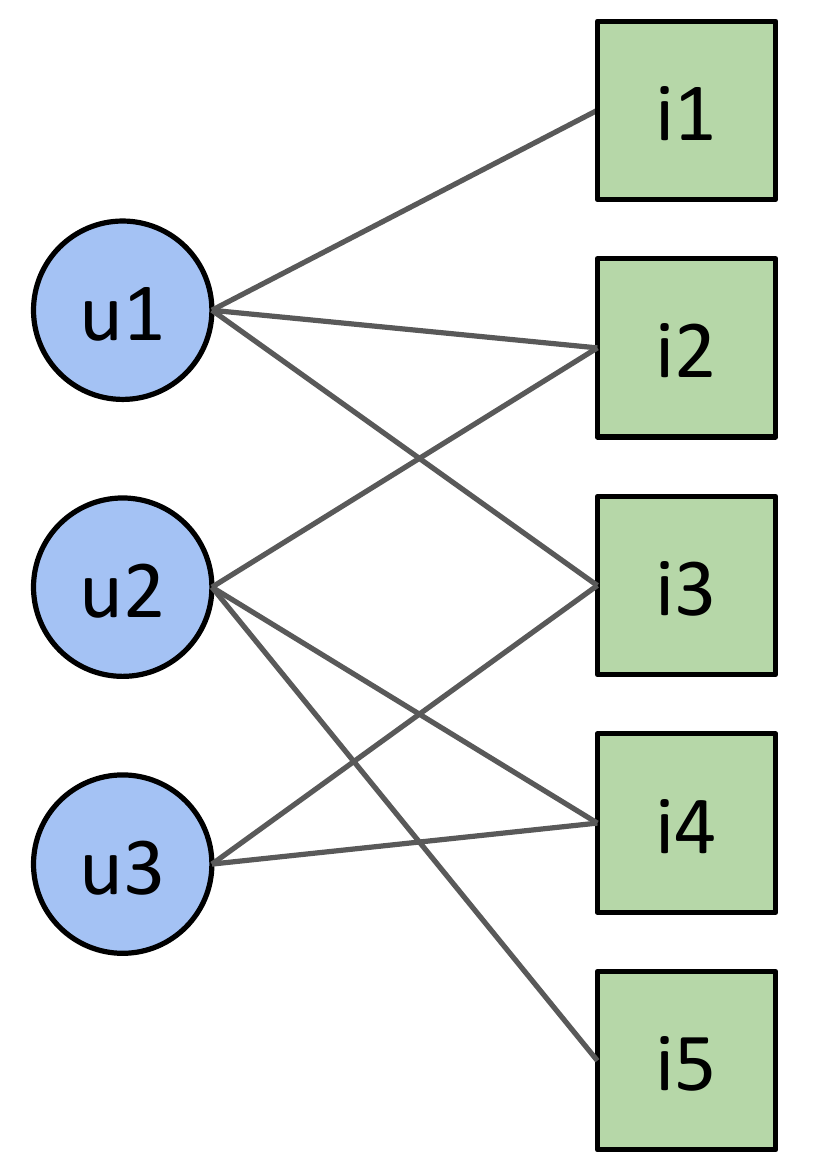}
\caption{}
\label{fig:mf-vs-gnn-1}
\end{subfigure}
\begin{subfigure}[b]{0.35\linewidth}
\includegraphics[width=0.95\linewidth]{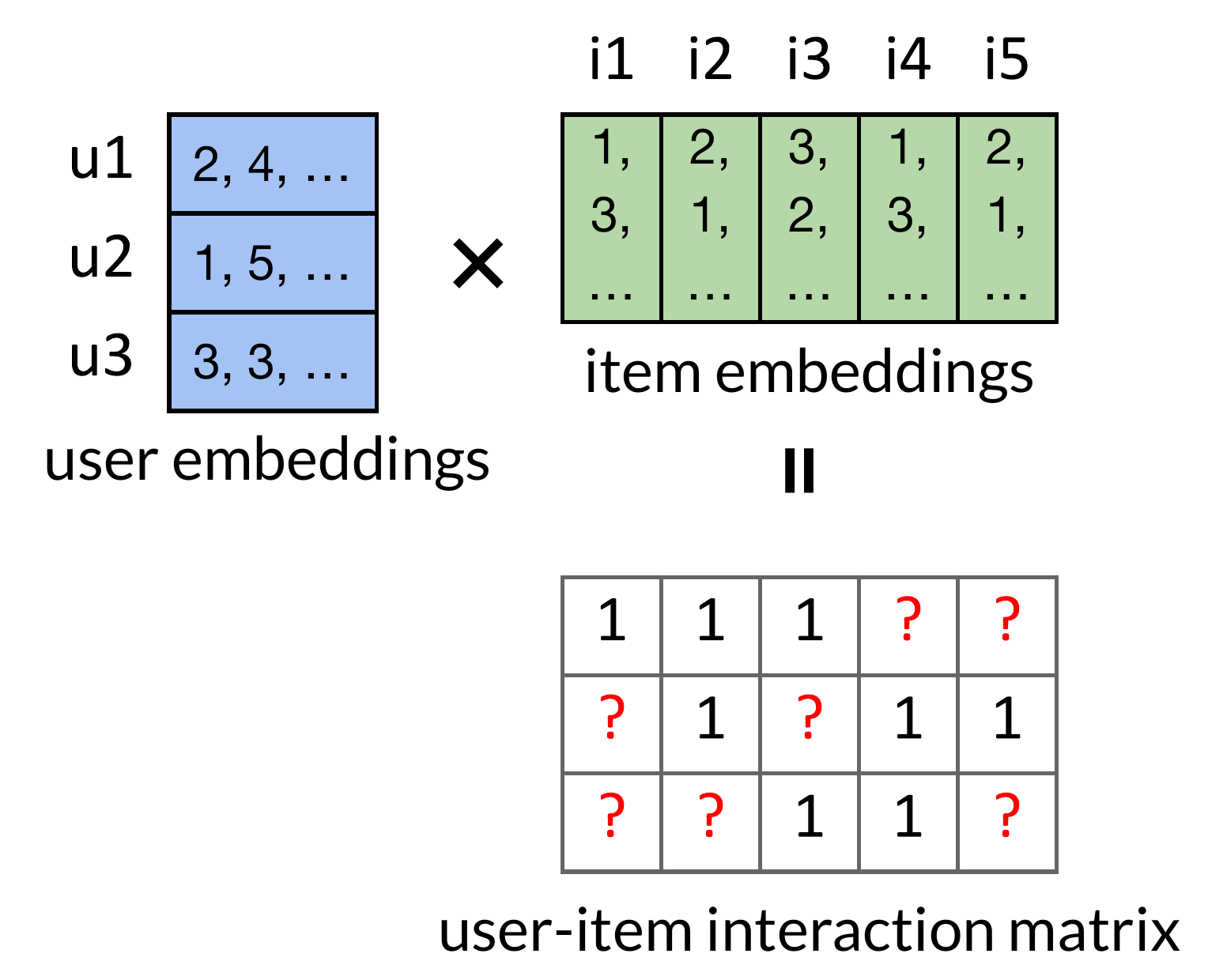}
\vspace{-0.5em}
\caption{}
\label{fig:mf-vs-gnn-2}
\end{subfigure}
\begin{subfigure}[b]{0.35\linewidth}
\includegraphics[width=1\linewidth]{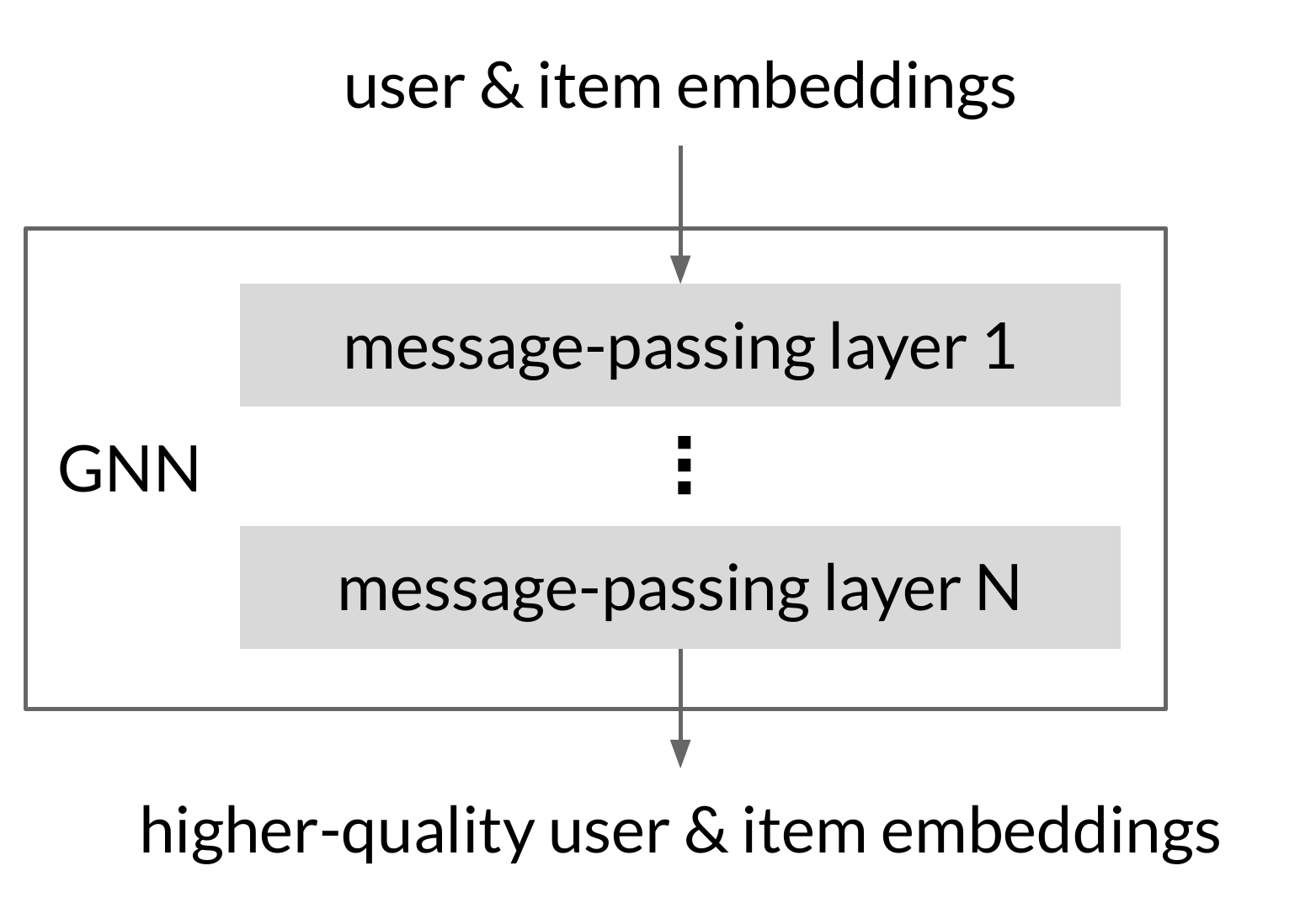}
\vspace{-1em}
\caption{}
\label{fig:mf-vs-gnn-3}
\end{subfigure}

\vspace{-1em}
\caption{Matrix factorization vs. GNNs for recommender systems --- (a) \textnormal{A user-item bipartite graph.} (b) \textnormal{Matrix factorization encodes each user/item into an embedding vector such that the dot product between the embeddings of a <user, item> pair can indicate the user’s preference for the item.} (c) \textnormal{GNNs improve upon matrix factorization by learning higher-quality embeddings through iterate message passing.}}
\label{fig:mf-vs-gnn}
\vspace{-1em}
\end{figure*}

Persistent memory technologies offer an alternative solution for large-scale GNNRecSys training by providing a significantly larger memory capacity than DRAMs.
For example, Intel Optane, which is a persistent memory product released in 2019, allows a single machine to have up to 6 TB of memory at an affordable cost, thus making single-machine full-graph GNN training feasible.
Single-machine full-graph GNN training eliminates the aforementioned inefficiencies in distributed subgraph GNN training, and also avoids substantial reworking of implementations needed by distributed processing.
One major concern of using persistent memory devices for GNNRecSys is their relatively low bandwidth compared with DRAMs.\footnote{Our measurement shows that the read bandwidth of Optane is 40\% that of DRAM and the write bandwidth of Optane is only 20\% that of DRAM. This result is consistent with prior studies \cite{yang2020empirical, izraelevitz2019basic}.}
This limitation can be particularly detrimental to achieving high performance for GNNRecSys workloads since their dominant compute kernels are sparse and memory access intensive.
Hence, this work seeks to better understand the following:

\smallskip
\emph{Can single-machine full-graph GNNRecSys training on persistent memory outperform distributed subgraph GNNRecSys training despite the relatively low memory bandwidth?}
\smallskip

To answer the question, we perform an in-depth characterization of GNNRecSys workloads and a comprehensive analysis of their performance on persistent memory.
We conduct experiments on twelve variants of two representative GNN models across a wide collection of datasets, taking into consideration different configurations of Optane.\footnote{We focus on the use of Optane as large volatile memory.}

Our analysis reveals that although the performance of GNNRecSys is negatively impacted by the relatively low bandwidth of Optane, using Optane together with DRAMs can largely recover the performance.
Furthermore, managing the hybrid Optane$+$DRAM memory system at a granularity of pages through the OS (in AppDirect Mode) works better than managing it at a granularity of cache lines by the hardware (in Memory Mode).
This is because GNNRecSys workloads have a large memory access size due to the embedding vectors.

Besides, this work provides guidance on how to configure Optane for GNNRecSys.
More concretely, both the memory consumption and execution time of GNNRecSys workloads are dominated by two types of sparse tensor compute kernels, namely, sampled dense-dense matrix multiplication (SDDMM) and sparse-dense matrix multiplication (SpMM).
These two types of compute kernels require different settings for optimal performance, including the NUMA data placement policy, the number of threads, and whether to use non-temporal write instructions.

In addition, we identify that a large batch size is crucial to realizing the advantages of full-graph GNNRecSys training, and manage to increase the batch size without incurring accuracy loss by linear learning rate scaling and a warm-up training strategy.
With the tuned batch size and optimal system configuration, Optane-based single-machine GNNRecSys training demonstrates a significant speedup over distributed subgraph training when handling deep GNN models (with at least two layers).
The speedup mainly comes from the ability of single-machine training to avoid cross-batch redundant computation that is pervasive in distributed training.



The rest of the paper is organized as follows.
Section \ref{sec:background} reviews the background of GNNRecSys, explains the inefficiencies of distributed subgraph GNN training, and introduces Optane.
Section \ref{sec:setup} describes the experiment setup.
Section \ref{sec:workload} presents workload characterization of the GNN models that we use for the experiments.
Section \ref{sec:optane-bandwidth} reports bandwidth measurements of Optane and discusses potential implications.
Section \ref{sec:kernel-level-benchmarking} and \ref{sec:end2end-benchmarking} present kernel-level and end-to-end results, respectively.
Section \ref{sec:discussion} discusses opportunities for further optimizations as well as how the findings of this work might generalize beyond Optane.
We survey related work in Section \ref{sec:related} and summarize in Section \ref{sec:conclusion}.

\section{Background}
\label{sec:background}

\subsection{GNN-Based Recommender Systems}
\label{subsec:gnn-recsys}

Recommender systems are crucial to the business of online services in the era of information explosion.
Authors in \cite{microsoft-recsys-report} reported that up to 75\% of movies watched on Netflix and 60\% of videos consumed on YouTube come from their recommender systems.


The conventional approach to building recommender systems is collaborative filtering through matrix factorization \cite{matrix-factorization-1, matrix-factorization-2}.
Figure \ref{fig:mf-vs-gnn-1} illustrates an example of a user-item bipartite graph where an edge can represent a click, a purchase, etc.
The matrix factorization method encodes each user/item into an embedding vector such that the dot product between the embeddings of a <user, item> pair can indicate the user's preference for the item; it updates the embeddings based on observed user-item interactions using gradient descent and then uses the obtained embeddings to predict missing interactions.

Recently, GNNs \cite{gcn, gat, graphsage} are gaining popularity as an effective method for learning high-quality vertex embeddings, thus offering a new approach to building recommender systems.
Unlike the matrix factorization method that directly feeds the user/item embeddings into the prediction function (e.g., dot product), GNNs refine the embeddings through \emph{message passing} on the user-item interaction graph.
More concretely, in the message passing paradigm, each vertex computes a new representation by aggregating messages from its incoming edges.
Let $G(V, E)$ denote a graph with a set of vertices $V$ and a set of edges $E$, $(src, e, dst)$ denote an edge $e$ pointing from vertex $src$ to $dst$, and $\mathbf{x}_v$ denote the initial embedding vector associated with vertex $v$.
The message passing paradigm carries out the following computations:

\vspace{-1em}

\begin{equation}\label{eq:message-generation}
\mathbf{m}_e = \phi(\mathbf{x}_{src}, \mathbf{x}_{dst}), (src, e, dst) \in E
\end{equation}

\vspace{-1em}

\begin{equation}\label{eq:message-aggregation}
\mathbf{h}_{dst} = \bigoplus\mathbf{m}_e, (src, e, dst) \in E
\end{equation}

\vspace{-1em}

\begin{equation}\label{eq:embed-transform}
\mathbf{x}_{dst}^{new} = \psi(\mathbf{h}_{dst})
\end{equation}


Here $\phi$, $\bigoplus$, and $\psi$ are customizable functions for generating messages, aggregating messages, and updating the embedding, respectively.
A GNN model iteratively applies Equations \eqref{eq:message-generation}~\eqref{eq:message-aggregation}~\eqref{eq:embed-transform} --- we call one such iteration a message-passing layer as shown in Figure \ref{fig:mf-vs-gnn-3} --- so that a vertex can incorporate its multi-hop neighborhood information into its embedding.
Prior studies \cite{ngcf, lightgcn} and our experiments show that increasing the number of message-passing layers improves the performance of GNNRecSys.\footnote{Matrix factorization can be viewed as a zero-layer GNN model.}
When auxiliary attributes are available, such as the user's age, gender, etc., GNNs can utilize them to augment the initial embeddings. We refer interested readers to \cite{gnn-recsys-survey}
for a comprehensive survey on the landscape of GNNRecSys research.

GNNRecSys workloads demand a large amount of memory.
More concretely, the memory consumption of Equation \eqref{eq:message-generation} is $len(\mathbf{m}) \times |E|$, of Equation \eqref{eq:message-aggregation} is $len(\mathbf{h}) \times |V|$, and of Equation \eqref{eq:embed-transform} is $len(\mathbf{x}) \times |V|$.
The above analysis only considers the forward propagation; training would double the memory consumption.
Our profiling shows that on a graph with one million vertices and three hundred million edges, training a three-layer GNN model with $len(\mathbf{m})$, $len(\mathbf{h})$, and $len(\mathbf{x})$ all set to 128 requires 500 GB of memory, easily exceeding the DRAM capacity on a typical server.
In comparison, running PageRank on the same graph only requires 3 GB of memory.

\subsection{Distributed Subgraph GNN Training}
\label{subsec:subgraph}

To tackle the memory capacity bottleneck in large-scale GNN training, existing efforts resort to distributed GNN training frameworks, such as DistGNN \cite{distgnn}, P3 \cite{p3}, and DistDGL \cite{distdgl}, among which only DistDGL is open-source.
DistDGL builds upon DGL \cite{dgl}, a widely-used GNN framework.

\begin{figure}[t]
\centering
\includegraphics[width=0.95\linewidth]{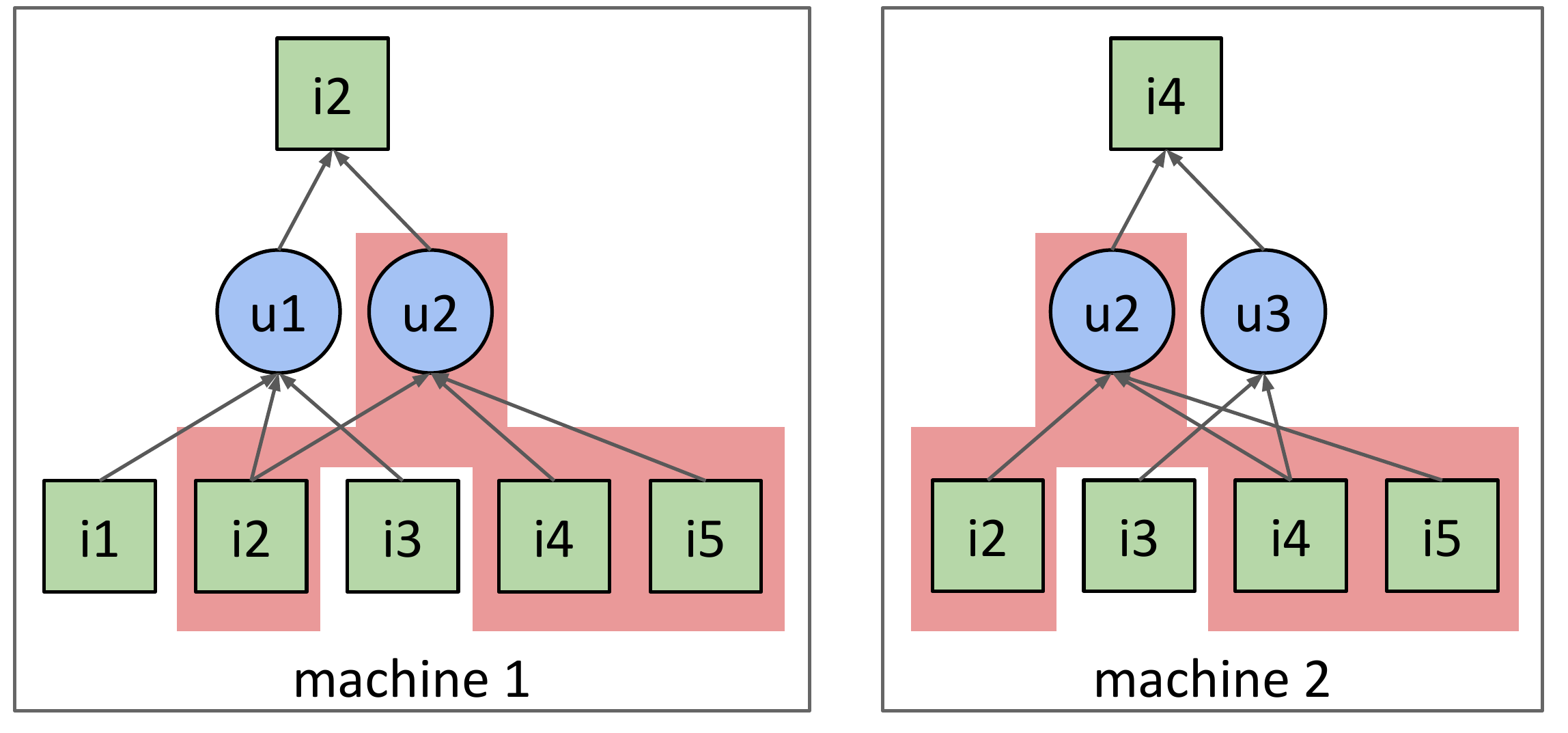}
\vspace{-1em}
\caption{Redundancy across subgraphs.}
\label{fig:subgraph-training}
\vspace{-1.5em}
\end{figure}

Figure \ref{fig:subgraph-training} illustrates the subgraph training approach adopted by DistDGL.
Let us assume the GNN model has two layers; there are two machines, and the batch size on each machine is one (the aggregate batch size is therefore two).
On machine 1, DistDGL selects $i2$ as the target vertex and constructs a subgraph consisting of $i2$ and its one- and two-hop neighbors --- we can apply two message-passing layers on the subgraph to compute the embedding for $i2$.
Similarly, on machine 2, DistDGL constructs a subgraph for $i4$.
One can immediately see the issue of redundancy, in both computation and memory consumption, across these two subgraphs.

In addition, because of the exponential growth of the subgraph size, only a limited batch size is allowed.
For faster training convergence, a large batch size is desired \cite{imagenet-1-hour, large-batch-bert} since fewer steps are needed to iterate the data points (i.e., one epoch).
One method to reduce the subgraph size is sampling \cite{graphsage, fastgcn, sampling-variance-reduction}, i.e., picking a small number of neighbors for each vertex instead of considering all the neighbors.
Sampling, however, incurs accuracy loss \cite{roc, dorylus}.
Besides, sampling is less effective in reducing the subgraph size for deep GNN models \cite{deepgcn, gnn-1000-layers}.

Furthermore, DistDGL suffers communication overhead.
DistDGL distributes the graph data across multiple machines, launches a subgraph builder (either with or without sampling) and a trainer on each machine, and performs synchronous training.
Inter-machine communication is needed for subgraph builders to fetch data located on a different machine, and for trainers to exchange gradients.

\subsection{Intel Optane Persistent Memory}
\label{subsec:optane}
The long-awaited persistent memory became commercially available with the release of Intel Optane in 2019.
Optane comes in the same form factor as DDR4 DRAM modules but with a higher memory density, allowing up to 6 TB of memory on a single machine.
This large capacity makes Optane attractive for data-intensive applications \cite{autotm, sampling-variance-reduction, optane-olap}.
Optane can be configured in the following two modes.

\textbf{Memory Mode.}
In this configuration, Optane is used as volatile main memory, and DRAM serves as a direct-mapped cache with a cache line size of 64 bytes.
The CPU's memory controller manages the cache transparently.
Software simply sees a large pool of volatile memory over which it has no explicit control.
This configuration allows the execution of existing code without any modifications.


\textbf{AppDirect Mode.}
In this configuration, Optane appears as a separate memory device.
There are two common ways to manage the hybrid Optane$+$DRAM memory system.
The first approach is to expose Optane as a NUMA node and rely on existing NUMA memory management utilities, e.g., \texttt{numactl}.
Here the memory management granularity is a page, i.e., 4 KB, instead of a cache line as in Memory Mode.
This approach requires no code modifications and at the same time gives programmers flexibility to try different NUMA memory management policies.
The second approach is to expose Optane as a file system.
Applications can then map files to memory to access them via load/store instructions.
This approach gives programmers explicit control over memory allocations, thus allowing for exploring application-specific memory management policies.
Specialized systems and libraries such as PMDK \cite{pmdk}, libmemkind \cite{libmemkind}, X-Mem \cite{x-mem}, and HeMem \cite{hemem} follow this approach.

\begin{table}
\caption{Different ways of using Optane.}
\label{tab:optane-configuration}
\vspace{-1em}
\centering
\begin{adjustbox}{width=0.95\linewidth}
\begin{tabular}{c|ccc}
\toprule
\multirow{2}{*}{} & \multirow{2}{*}{Memory Mode} &\multicolumn{2}{c}{AppDirect Mode} \\
\cmidrule{3-4}
& & NUMA & file system \\
\midrule
managed by         & hardware    & OS      & programmer \\
granularity        & cache line  & page    & arbitrary size \\
flexibility        & low         & medium  & high \\
code modification? & no          & no      & yes \\
\bottomrule
\end{tabular}
\end{adjustbox}
\vspace{-1em}
\end{table}

Table \ref{tab:optane-configuration} summarizes the different ways of using Optane. This work focuses on the first two, i.e., using Optane in Memory Mode and in AppDirect Mode through NUMA utilities.
We discuss potential performance gains by specializing the memory management policy for GNNRecSys workloads and leave it for future research.

\section{Experiment Setup}
\label{sec:setup}

\textbf{Datasets.}
Table \ref{tab:datasets} lists the datasets that we use for the experiments.
\texttt{movielens-10m} \cite{movielens}, \texttt{gowalla} \cite{gowalla}, and \texttt{amazon-book} \cite{amazon-book} are real-world datasets for recommending movies, locations, and books, respectively.
They have been widely used in GNNRecSys research \cite{gcmc, ngcf, lightgcn}.
To work around the issue that industry-scale datasets for recommender systems are not publicly available, we follow the Kronecker expansion method proposed in \cite{kronecker-expansion} to synthetically generate large datasets by expanding existing small ones.
Kronecker expansion can preserve the original graph's characteristics such as power-law degree distribution, community structure, item popularity, etc.
We expand \texttt{movielens-10m}, \texttt{gowalla}, and \texttt{amazon-book} to three medium-size graphs (around 300M edges) and three large-size graphs (around one billion edges); \texttt{m-x25} denotes a graph that is expanded from \texttt{movielens-10m} by a factor of 25 in terms of the number of edges.
We use \texttt{movielens-10m}, \texttt{gowalla}, and \texttt{amazon-book} to verify the GNN model's accuracy, and the six synthetic datasets for performance benchmarking.

\begin{table}[t]
\caption{Datasets.}
\label{tab:datasets}
\vspace{-1em}
\centering
\begin{adjustbox}{width=0.95\linewidth}
\begin{tabular}{c|cccc}
\toprule
dataset & \# users & \# items & \# interactions & density \\
\midrule
\texttt{movielens-10m}   & 70K    & 11K     & 10M   & 1.34\% \\
\texttt{gowalla}         & 30K    & 41K     & 1M    & 0.08\% \\
\texttt{amazon-book}     & 53K    & 92K     & 3M    & 0.06\% \\
\midrule
\texttt{m-x25}           & 349K   & 53K     & 250M  & 1.34\% \\
\texttt{g-x256}          & 478K   & 656K    & 263M  & 0.08\% \\
\texttt{a-x100}          & 526K   & 916K    & 298M  & 0.06\% \\
\midrule
\texttt{m-x100}          & 699K   & 107K    & 1000M  & 1.34\% \\
\texttt{g-x1024}         & 955K   & 1311K   & 1052M  & 0.08\% \\
\texttt{a-x400}          & 1053K  & 1832K   & 1194M  & 0.06\% \\
\bottomrule
\end{tabular}
\end{adjustbox}
\vspace{-1em}
\end{table}

\begin{figure}[t]
\centering
\includegraphics[width=0.95\linewidth]{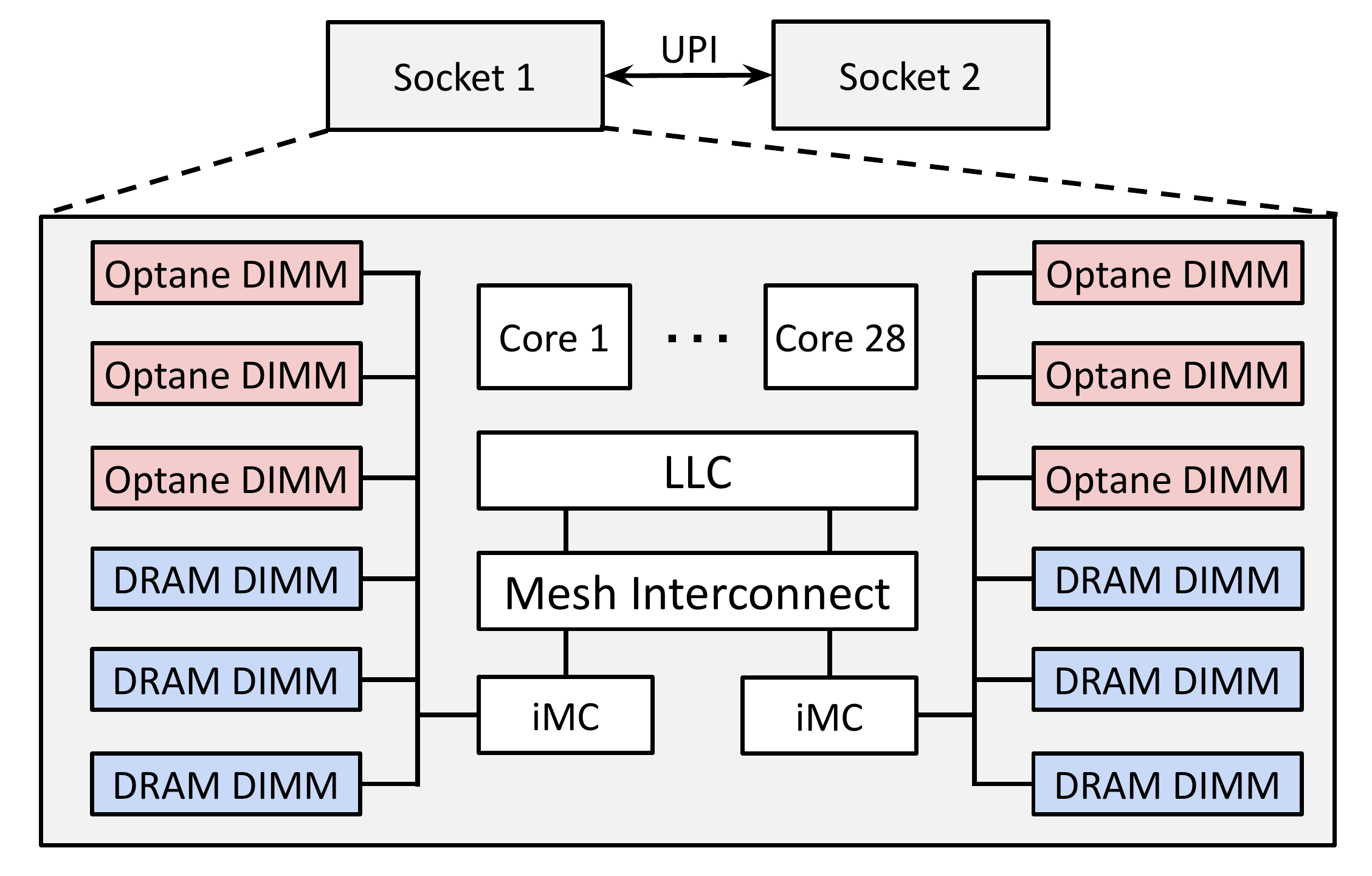}
\vspace{-1em}
\caption{A two-socket Intel Cascade Lake machine.}
\label{fig:optane-machine}
\vspace{-1em}
\end{figure}

\textbf{Optane Machine.}
Figure \ref{fig:optane-machine} depicts the architecture of the Optane machine.
It is a 2.2 GHz Intel Xeon Platinum 8276L machine with two sockets connected by the Intel Ultra Path Interconnect (UPI).
Each socket has 28 physical cores, two integrated memory controllers (iMCs), six DRAM DIMMs, and six Optane DIMMs.
Each DRAM DIMM is 32 GB; each Optane DIMM is 128 GB.
The total DRAM capacity is 384 GB ($32 \times 6 \times 2$); the total Optane capacity is 1536 GB ($128\times 6 \times 2$).

\textbf{Software.}
We use DGL v0.7.2 with PyTorch v1.7 backend to implement GNN models.
Transparent Huge Pages (THP) are enabled (default in Linux).
We use PyTorch Profiler\footnote{https://pytorch.org/tutorials/recipes/recipes/profiler\_recipe.html} to get execution time breakdown and \texttt{mprof}\footnote{https://github.com/pythonprofilers/memory\_profiler} to measure memory consumption.

\section{Workload Characterization}
\label{sec:workload}

We choose NGCF \cite{ngcf} and LightGCN \cite{lightgcn} for the experiments, both of which are representative GNN models that have been widely used in recommender systems.
Both NGCF and LightGCN follow the message-passing paradigm described in Section \ref{subsec:gnn-recsys}.
Compared with NGCF, LightGCN is less compute intensive due to a simplified architecture.

\begin{figure*}[t]
\centering

\begin{subfigure}[b]{0.32\linewidth}
\includegraphics[width=1\linewidth]{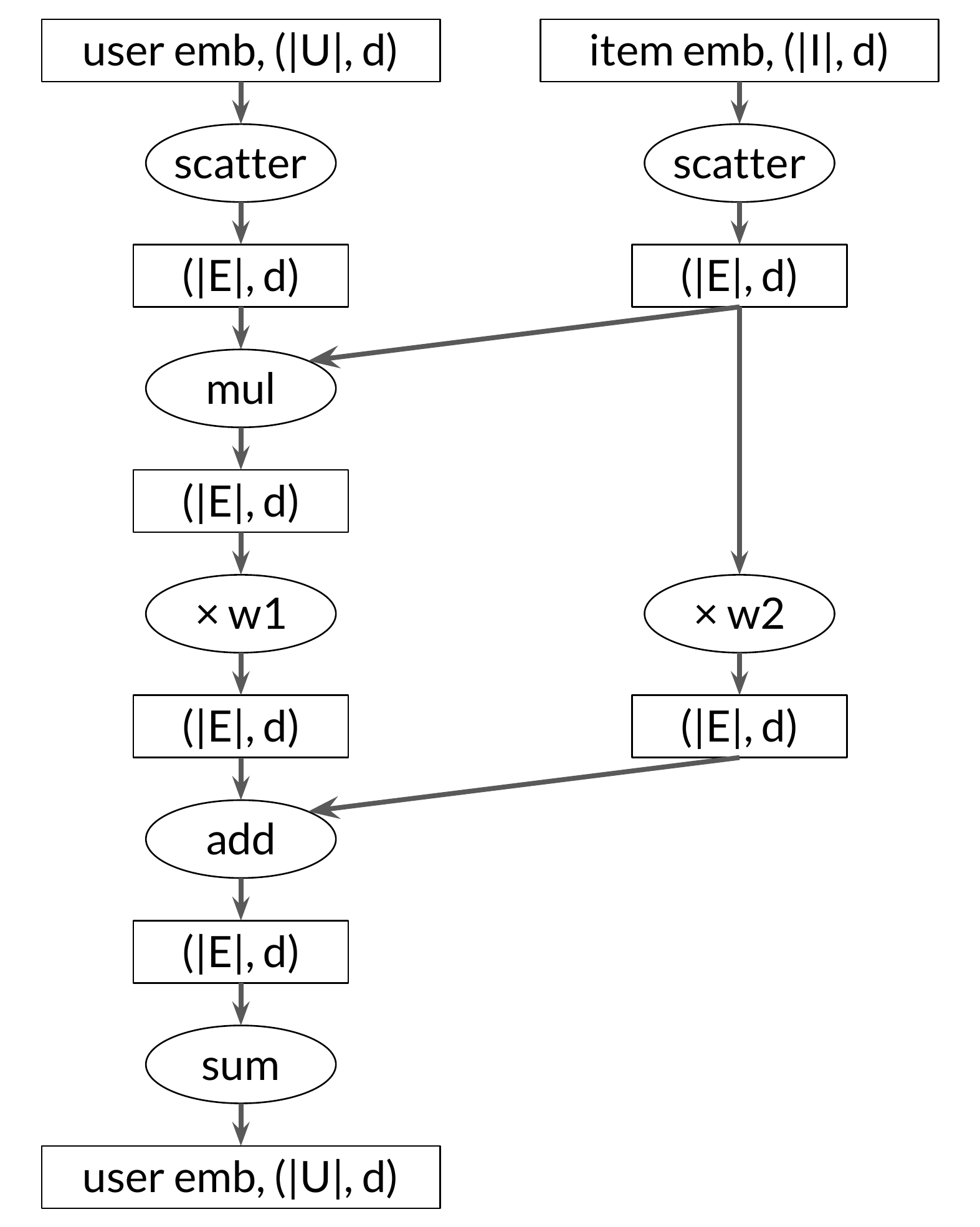}
\caption{}
\label{fig:ngcf-opt-1}
\end{subfigure}
\begin{subfigure}[b]{0.32\linewidth}
\includegraphics[width=1\linewidth]{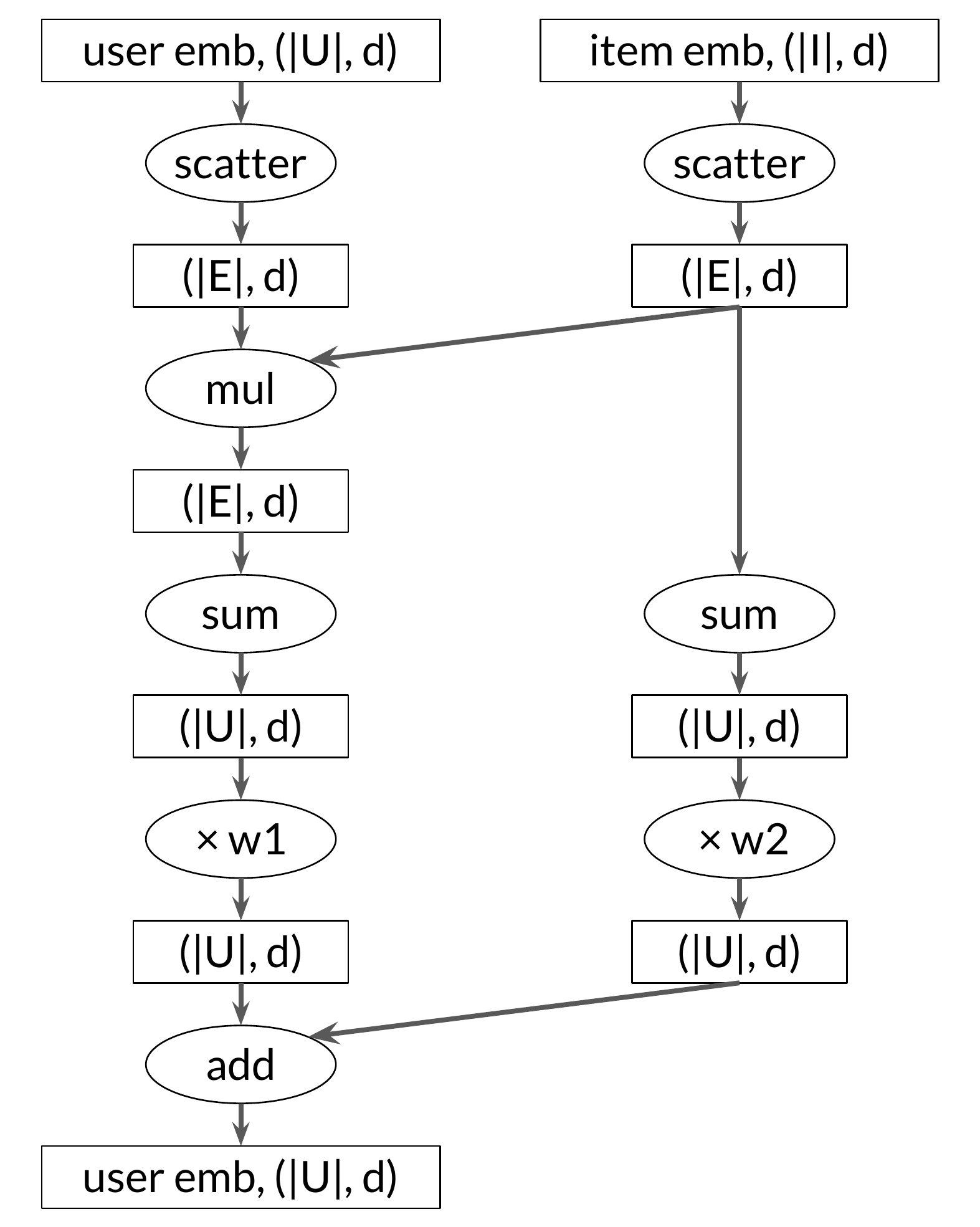}
\caption{}
\label{fig:ngcf-opt-2}
\end{subfigure}
\begin{subfigure}[b]{0.32\linewidth}
\includegraphics[width=1\linewidth]{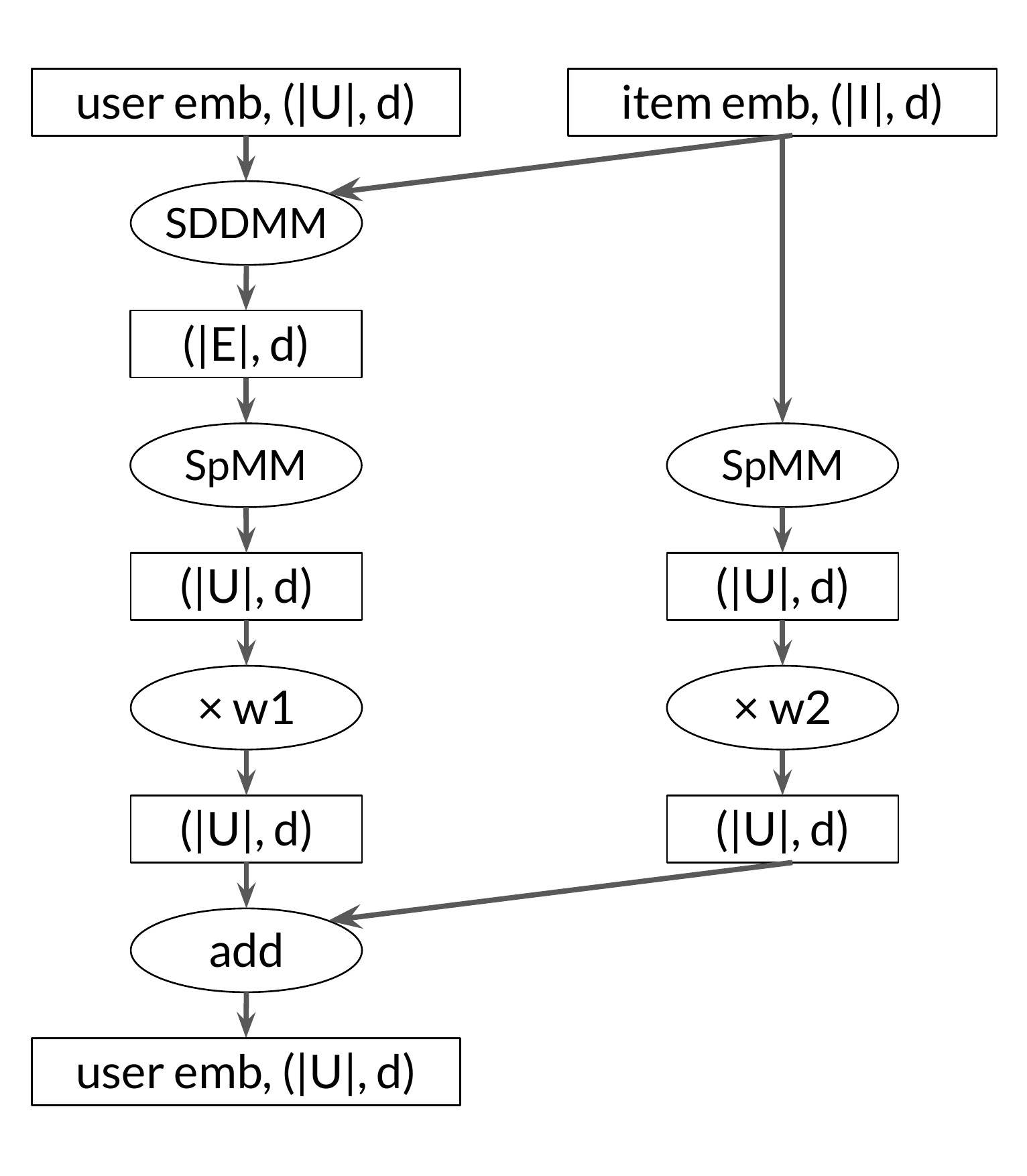}
\caption{}
\label{fig:ngcf-opt-3}
\end{subfigure}

\vspace{-1em}
\caption{NGCF optimizations --- (a) \textnormal{The original implementation in DGL.} (b) \textnormal{Reducing the computational complexity of multiplying weight matrices from $O(|E|)$ to $O(|U|)$ by switching the execution order.} (c) \textnormal{Replacing less-efficient scatter and sum operations with highly-optimized tensor compute kernels, i.e., SDDMM and SpMM.} \textnormal{The last optimization (reusing the SDDMM results) is not shown.}}
\label{fig:ngcf-opt}
\vspace{-1em}
\end{figure*}

\subsection{NGCF}

NGCF's message generation function, message aggregation function, and embedding update function are defined as follows. Here $\odot$ denotes element-wise multiply and $W1$ and $W2$ are trainable weight matrices.

\vspace{-1.5em}

\begin{equation}\label{eq:ngcf-message-generation}
\mathbf{m}_e = (\mathbf{x}_{src} \odot \mathbf{x}_{dst}) \times W1 + \mathbf{x}_{src} \times W2, (src, e, dst) \in E
\end{equation}

\vspace{-1.5em}

\begin{equation}\label{eq:ngcf-message-aggregation}
\mathbf{h}_{dst} = sum(\mathbf{m}_e), (src, e, dst) \in E
\end{equation}

\vspace{-1.5em}

\begin{equation}\label{eq:ngcf-embed-transform}
\mathbf{x}_{dst}^{new} = \mathbf{h}_{dst}
\end{equation}


We care about two parameters in the architecture: the embedding length $len(\mathbf{x})$, and the number of message-passing layers.
Increasing either one of these two parameters typically improves the model accuracy, but at the cost of higher memory consumption and computational complexity.
Specifically, for full-graph training, the memory consumption and computational complexity grow linearly with both the embedding length and the number of layers; for subgraph training, the growth is linear with the embedding length but exponential with the number of layers.
In the experiments, we vary these two parameters --- setting the number of layers to one or two or three, setting the embedding length to 128 or 256 --- to create six variants of NGCF.
We use NGCF-1L-128E to denote an NGCF model with one layer and the embedding length set to 128.

We recognize several inefficiencies in the original implementation of NGCF in DGL and apply a set of optimizations.
Figure \ref{fig:ngcf-opt-1} illustrates the dataflow graph (one message-passing layer) of the original implementation for computing user embeddings; the dataflow graph for computing item embeddings is symmetric.
Our first optimization is to switch the execution order between multiplying weight matrices and aggregating messages, that is, we convert $sum((\mathbf{x}_{src} \odot \mathbf{x}_{dst}) \times W1 + \mathbf{x}_{src} \times W2)$ into $sum(\mathbf{x}_{src} \odot \mathbf{x}_{dst}) \times W1 + sum(\mathbf{x}_{src}) \times W2$, as illustrated in Figure \ref{fig:ngcf-opt-2}.
This optimization reduces the computational complexity of multiplying weight matrices from $O(|E|)$ to $O(|U|)$, where $|E|$ is the number of edges and $|U|$ is the number of users.
Our second optimization is to replace less-efficient scatter and sum operations with sparse tensor compute kernels that have been highly optimized by the DGL framework.
Concretely, we use generalized SDDMM (sampled dense-dense matrix multiplication) and SpMM (sparse-dense matrix multiplication) to implement message generation and aggregation, respectively, as illustrated in Figure \ref{fig:ngcf-opt-3}.
While the standard SDDMM performs dot product in its inner most loop, a generalized one supports element-wise add, which is required by NGCF.
For SpMM, while the standard one performs add reduction, a generalized one supports max reduction among others.\footnote{https://docs.dgl.ai/api/python/dgl.ops.html}
Moreover, the gradient calculation of (generalized) SDDMM and SpMM are also mapped to these two kernels.
We refer interested readers to \cite{featgraph} for a thorough description on the connection between message passing and SDDMM/SpMM kernels.
Our last optimization is to reuse the SDDMM results obtained during computing user embeddings to compute item embeddings (since ${\mathbf{x}_{src} \odot \mathbf{x}_{dst} = \mathbf{x}_{dst} \odot \mathbf{x}_{src}}$), instead of calculating SDDMM twice as in the original implementation.

\begin{figure}[t]
\centering
\includegraphics[width=0.8\linewidth]{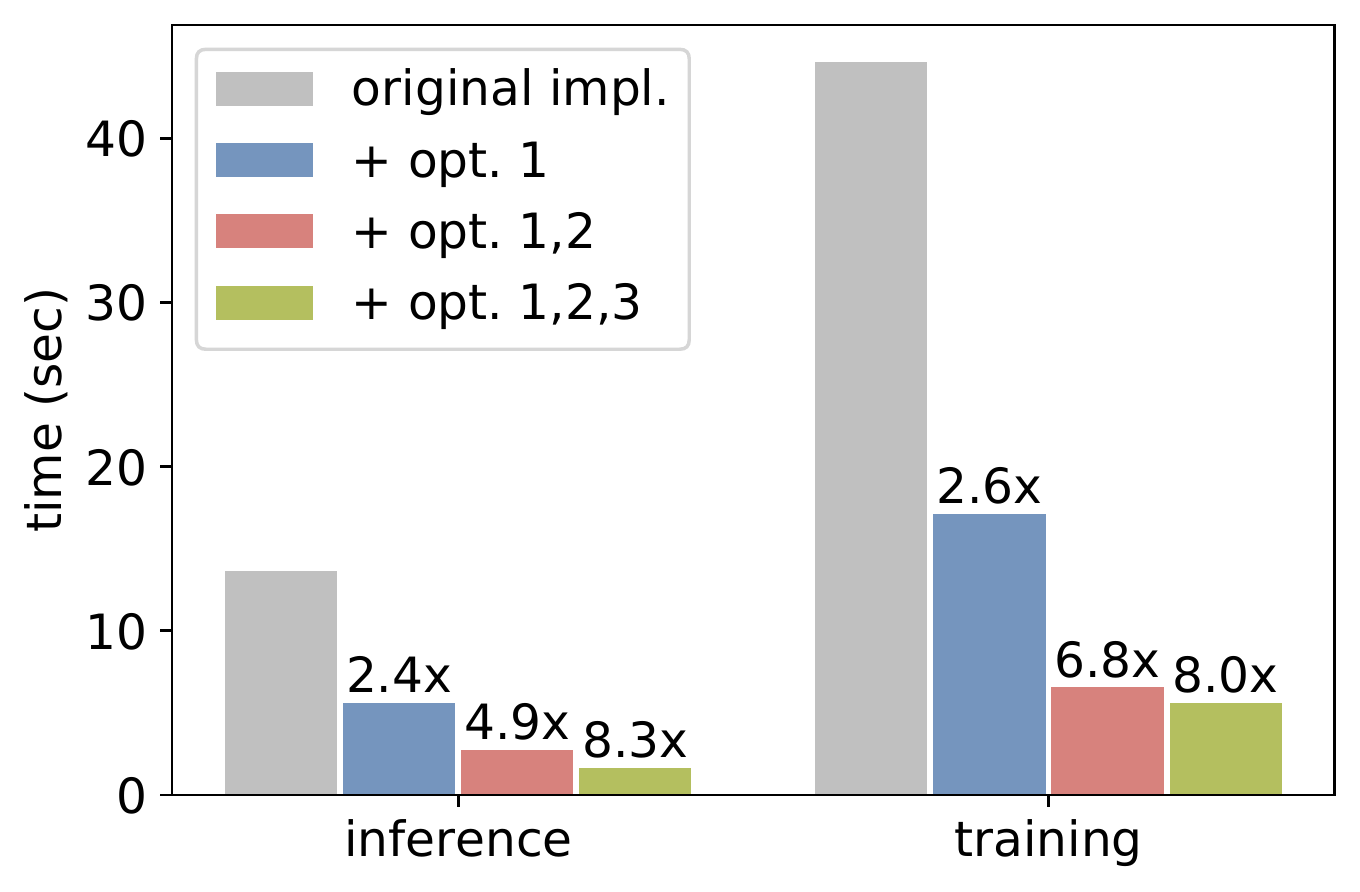}
\vspace{-1em}
\caption{Effects of NGCF optimizations --- \textnormal{The dataset is \texttt{movielens-10m}; the model is NGCF-3L-128E; experiments run without using Optane; time is for one epoch.}}
\label{fig:ngcf-opt-effects}
\end{figure}

\begin{figure}[t]
\centering
\includegraphics[width=0.95\linewidth]{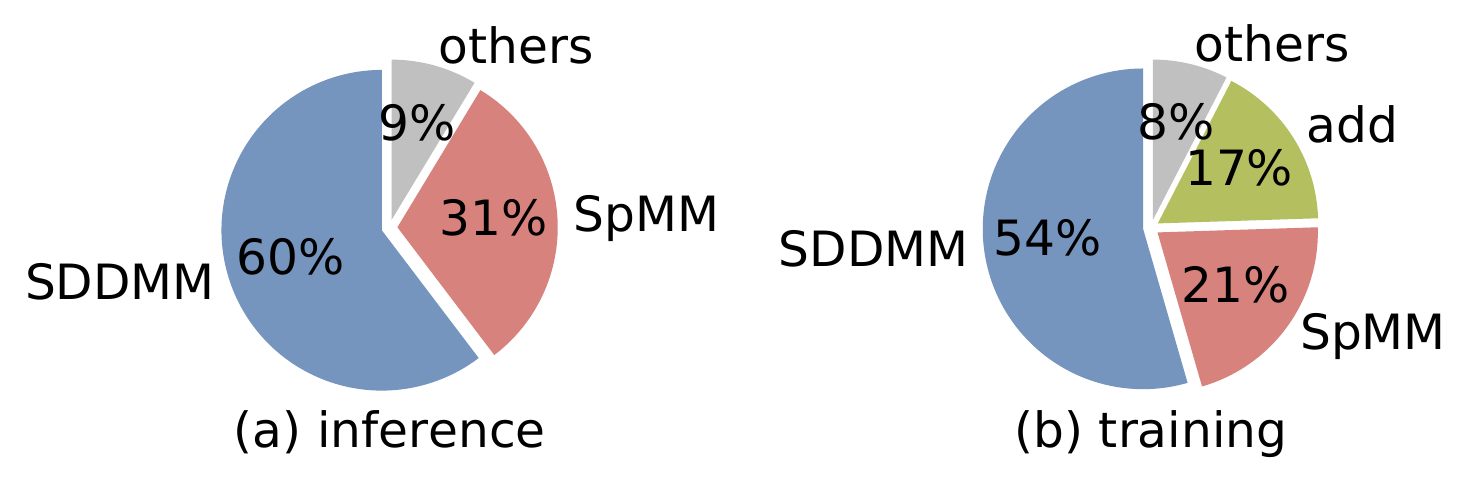}
\vspace{-1em}
\caption{Execution time breakdown of the optimized NGCF.}
\label{fig:ngcf-time-breakdown}
\vspace{-1em}
\end{figure}

Figure \ref{fig:ngcf-opt-effects} shows that combining the three optimizations accelerates the inference of NGCF-3L-128E by 8.3$\times$ and training by 8.0$\times$ on the \texttt{movielens-10m} dataset.
We will use the optimized implementation for performance analysis on Optane.
Figure \ref{fig:ngcf-time-breakdown} shows the execution time breakdown.
SDDMM and SpMM together take 91\% of the total time for inference and 75\% for training.
Besides, the add operation takes 17\% of the training time, which is mainly used for updating trainable weights with gradients during backpropagation.

\subsection{LightGCN}

LightGCN follows the overall architecture of NGCF but simplifies message generation by removing $W1$ and $W2$ in Equation \eqref{eq:ngcf-message-generation}.
The second and third optimizations presented in the previous subsection are also applied to LightGCN.
Our experiment results show that LightGCN is 10--30\% faster than NGCF (see Section \ref{sec:end2end-benchmarking}).
Furthermore, LightGCN spends a larger fraction of execution time on SDDMM and SpMM --- up to 95\% for inference and 85\% for training --- because it gets rid of multiplication with $W1$ and $W2$.


\section{Optane Bandwidth Measurements}
\label{sec:optane-bandwidth}

\begin{figure*}[t]
\centering

\begin{subfigure}[b]{0.33\linewidth}
\includegraphics[width=1\linewidth]{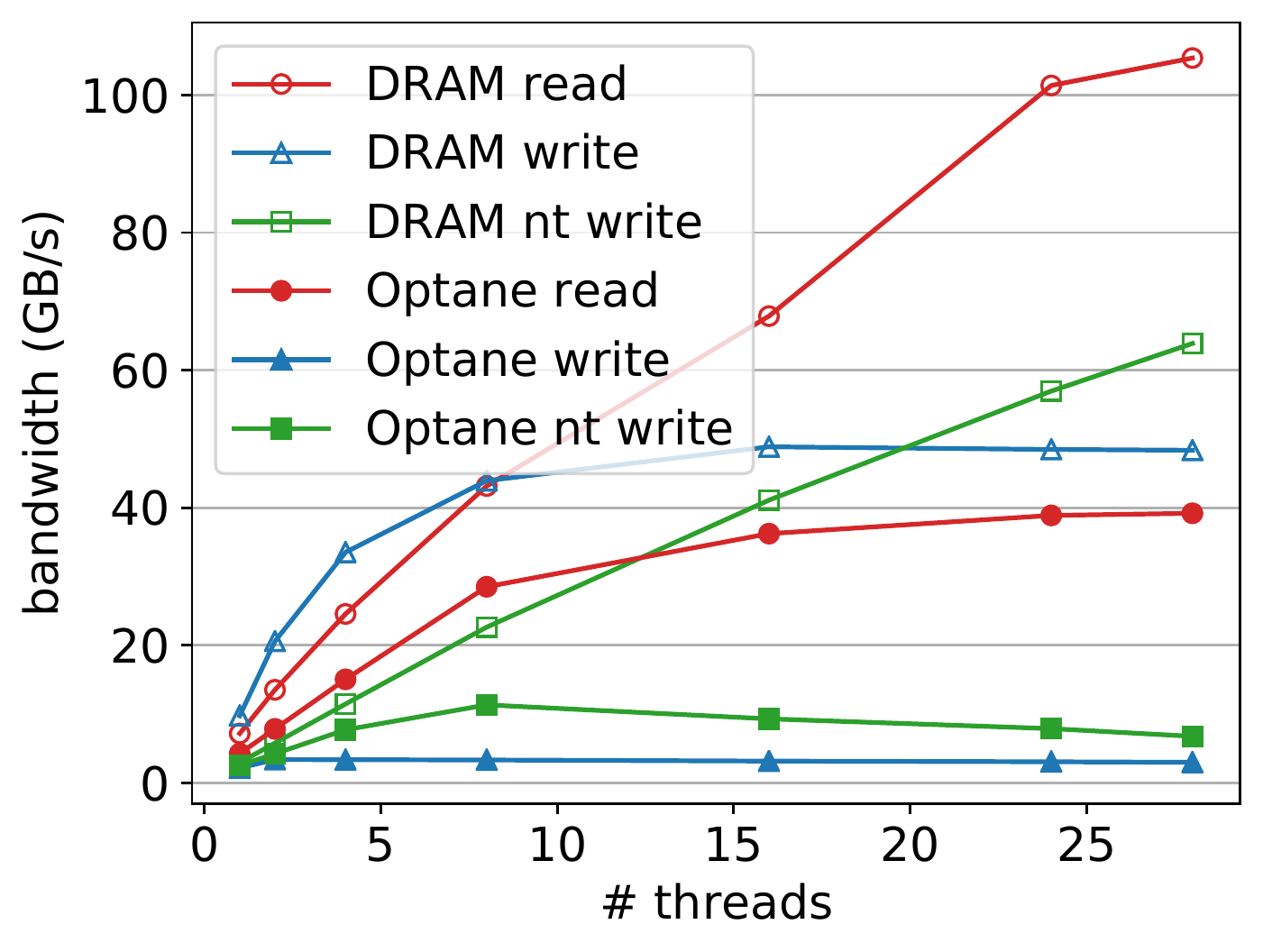}
\caption{\textnormal{Sequential memory accesses}}
\label{fig:bandwidth-1}
\end{subfigure}
\begin{subfigure}[b]{0.33\linewidth}
\includegraphics[width=1\linewidth]{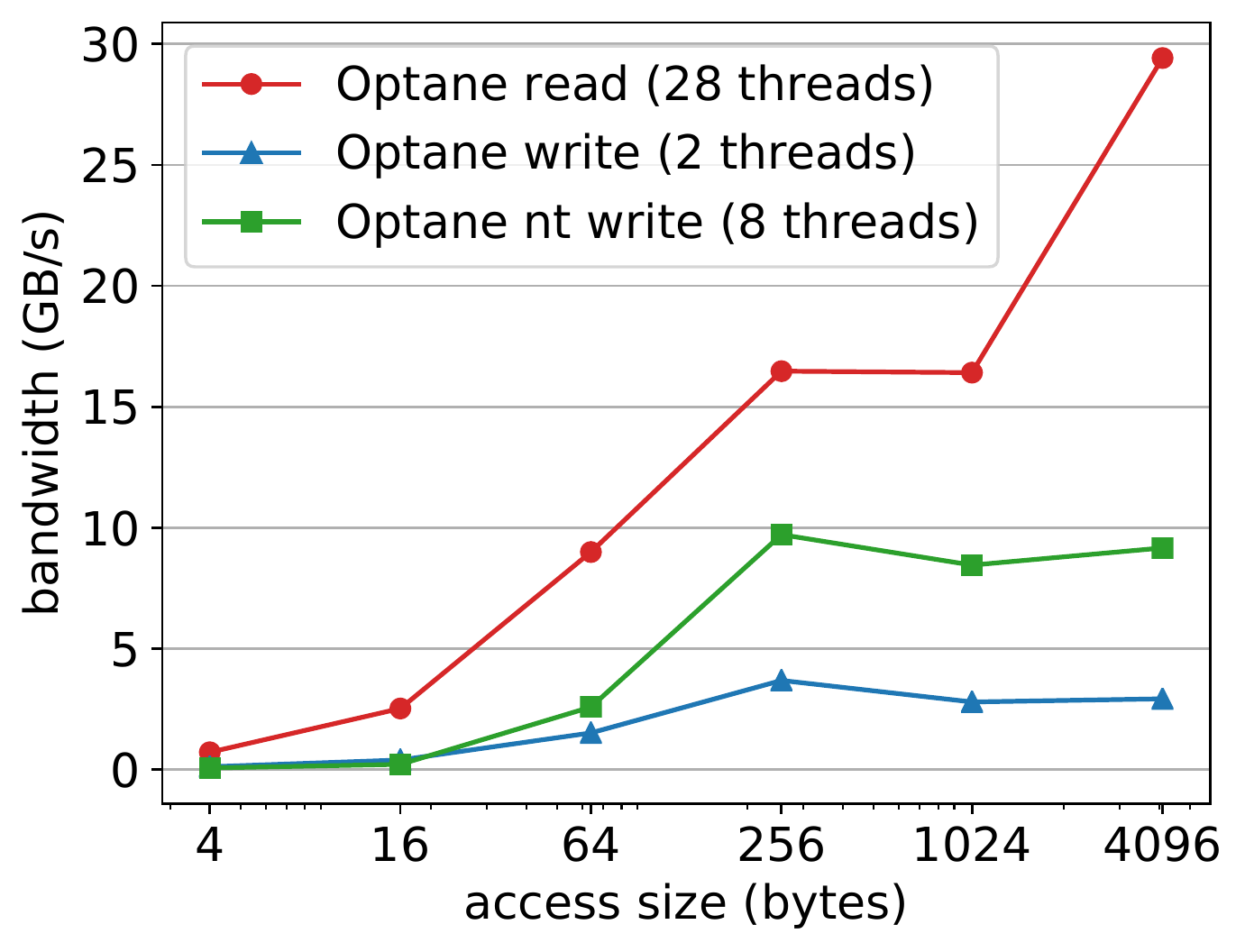}
\caption{\textnormal{Random memory accesses}}
\label{fig:bandwidth-2}
\end{subfigure}
\begin{subfigure}[b]{0.33\linewidth}
\includegraphics[width=1\linewidth]{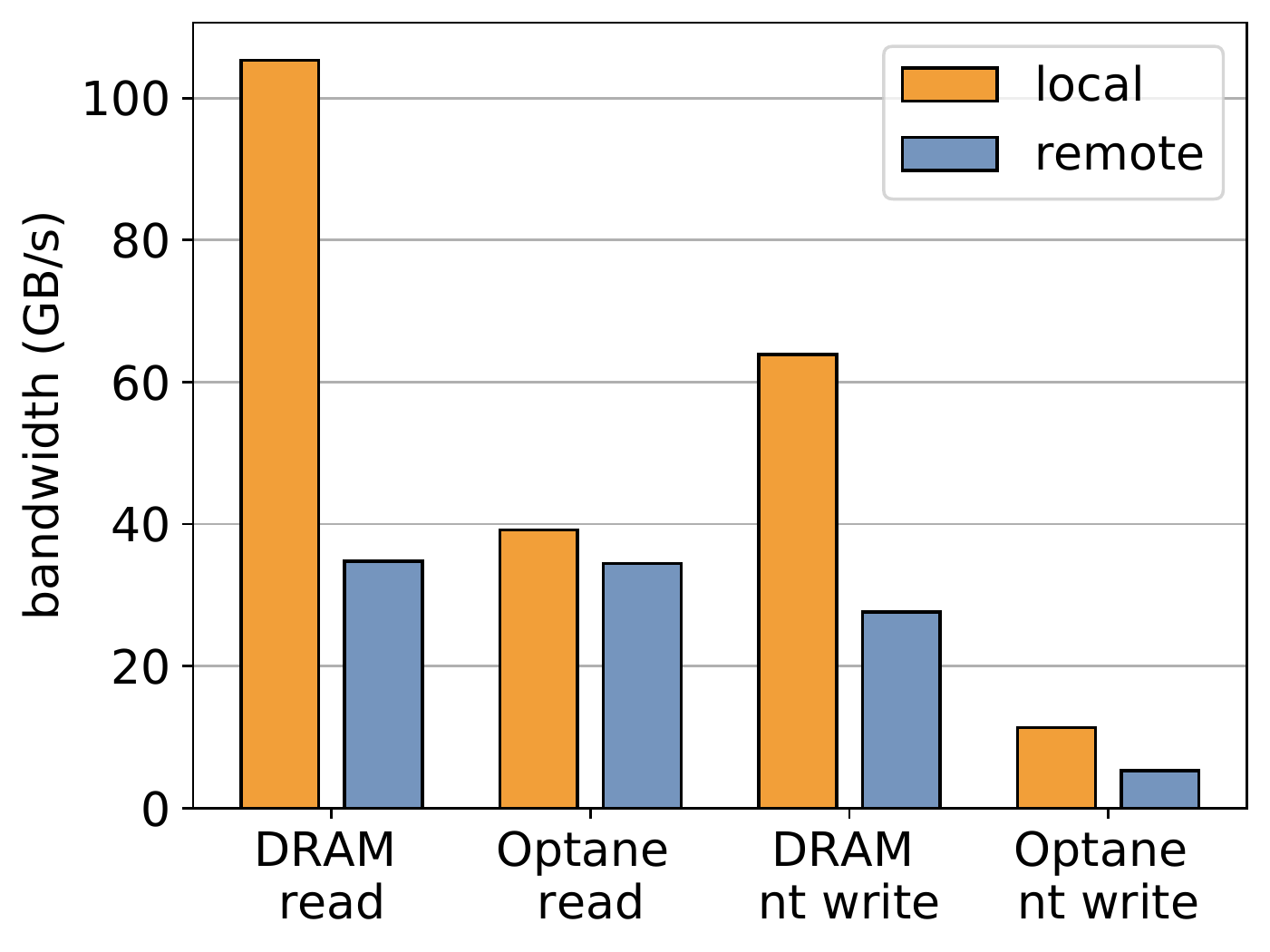}
\caption{\textnormal{NUMA effects}}
\label{fig:bandwidth-3}
\end{subfigure}

\vspace{-1em}
\caption{Bandwidth measurements.}
\label{fig:bandwidth}
\end{figure*}

Since GNNRecSys workloads are memory bandwidth bound and less sensitive to the memory latency \cite{gnnadvisor, p3}, we focus on bandwidth measurements.
Specifically, we measure the bandwidth of Optane under both sequential and random memory access patterns for both remote (i.e., cross-socket) and local accesses.

Figure \ref{fig:bandwidth-1} shows that for sequential memory accesses, the peak read bandwidth of Optane (with 28 threads) is 39.2 GB/s, which is 37\% of what DRAM offers; the peak write bandwidth of Optane (with 2 threads) is 3.5 GB/s, which is 7\% of that of DRAM; when using non-temporal write (nt write), the peak write bandwidth of Optane (with 8 threads) is 11.3 GB/s, which is 18\% of that of DRAM.
Compared to a normal write, an nt write achieves a higher bandwidth by bypassing the cache hierarchy and thus preventing the cache eviction mechanism from converting sequential stores into random stores \cite{sampling-variance-reduction}.

Figure \ref{fig:bandwidth-2} shows that for random memory accesses, the bandwidth utilization of Optane increases as the access size increases.
Specifically, the write bandwidth utilization saturates at an access size of 256 bytes; the read bandwidth utilization is not saturated even at 4096 bytes.
For GNNRecSys, the access size is typically hundreds of bytes depending on the embedding length, whereas for traditional graph processing workloads such as PageRank, it is only four bytes.

Figure \ref{fig:bandwidth-3} shows that the remote read bandwidth of Optane is comparable to that of DRAM (34.5 vs. 35.8 GB/s); both are limited by the UPI bandwidth.
The remote nt write bandwidth of DRAM is even higher than the local nt write bandwidth of Optane, indicating that it could be beneficial to allocate memory on all the sockets to utilize more DRAMs even when the data can fit into Optane on a single socket.

To understand the implications of Optane on the performance of GNNRecSys workloads, we have the following questions to be answered:


\begin{tightlist}{1.5em}{0.5em}{0.5em}
    \item [1.] To what extent does the relatively low memory bandwidth (especially write) of Optane hurt the performance of GNNRecSys workloads?
    \item [2.] Is it beneficial to use nt write?
    \item [3.] What is the optimal NUMA configuration?
    \item [4.] What is the tradeoff between accuracy and efficiency with different embedding lengths?
\end{tightlist}

\section{Kernel-Level Analysis and Optimization}
\label{sec:kernel-level-benchmarking}

In this section, we aim to (1) understand the implications of Optane on the performance of SDDMM and SpMM, the two dominant compute kernels in NGCF and LightGCN, and (2) give suggestions on how to configure Optane to achieve high performance.

\begin{figure}[t]
\centering
\includegraphics[width=1\linewidth]{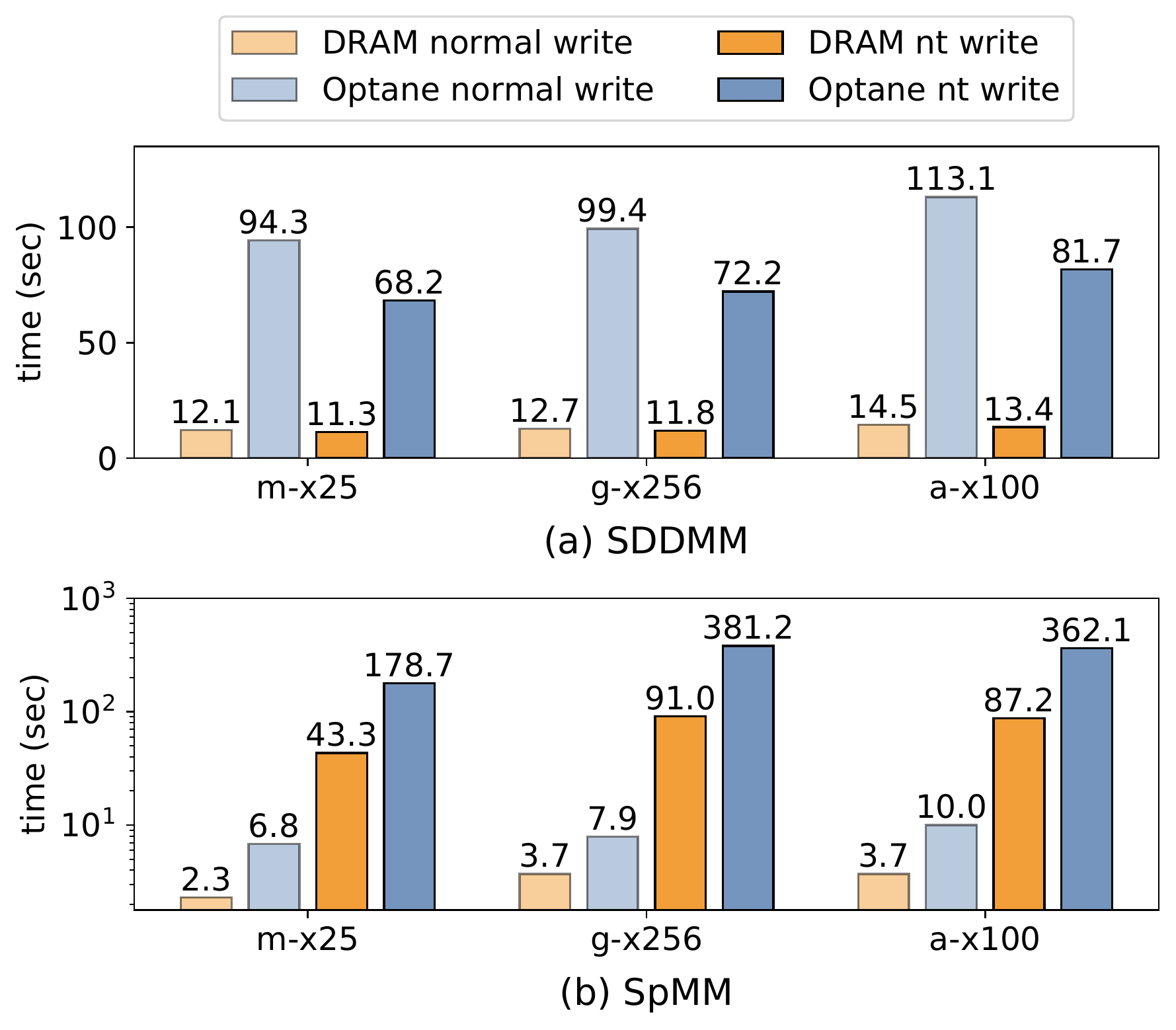}
\vspace{-2em}
\caption{Execution time of SDDMM and SpMM using Optane vs. DRAM --- \textnormal{The embedding length is 128; experiments run on a single socket using 28 threads; data fit into DRAM.}}
\label{fig:kernel-perf-optane-vs-dram}
\vspace{-1em}
\end{figure}

Figure \ref{fig:kernel-perf-optane-vs-dram} shows the execution time of SDDMM and SpMM using Optane vs. DRAM, tested on the three synthetic million-edge graphs, i.e., \texttt{m-x25}, \texttt{g-x256}, and \texttt{a-x100}.
Using Optane alone gives a lower bound for the achievable performance since in practice we always use Optane together with DRAM.
The main results are:
(1) When using normal write, the performance of SDDMM on Optane is 7.7--7.8$\times$ lower than that on DRAM; for SpMM, this number is 2.2--3.0$\times$.
SDDMM has a severe slowdown on Optane because SDDMM is write intensive and the write bandwidth of Optane is less than 10\% that of DRAM (see Figure \ref{fig:bandwidth-1}).
In comparison, SpMM is read intensive and therefore sees a more modest slowdown on Optane.
(2) When switching from normal write to nt write, the performance of SDDMM improves by 1.1$\times$ on DRAM and 1.4$\times$ on Optane; the performance of SpMM, however, degrades significantly --- more than 20$\times$ for both DRAM and Optane.
This is because SpMM has a large degree of temporal locality due to its aggregation computation pattern; hence, using nt write that bypasses the cache hierarchy only hurts its performance.
In comparison, SDDMM has no temporal locality in accessing its outputs and therefore can benefit from the higher write bandwidth brought by nt write.
(3) Among the three datasets, which have a similar number of edges, SpMM runs fastest on \texttt{m-x25} for both Optane and DRAM.
This is because \texttt{m-x25} has a higher density, leading to a higher degree of locality and consequently higher cache utilization.
In remaining experiments, we use nt write for SDDMM and normal write for SpMM.

\begin{figure}[t]
\centering
\includegraphics[width=1.0\linewidth]{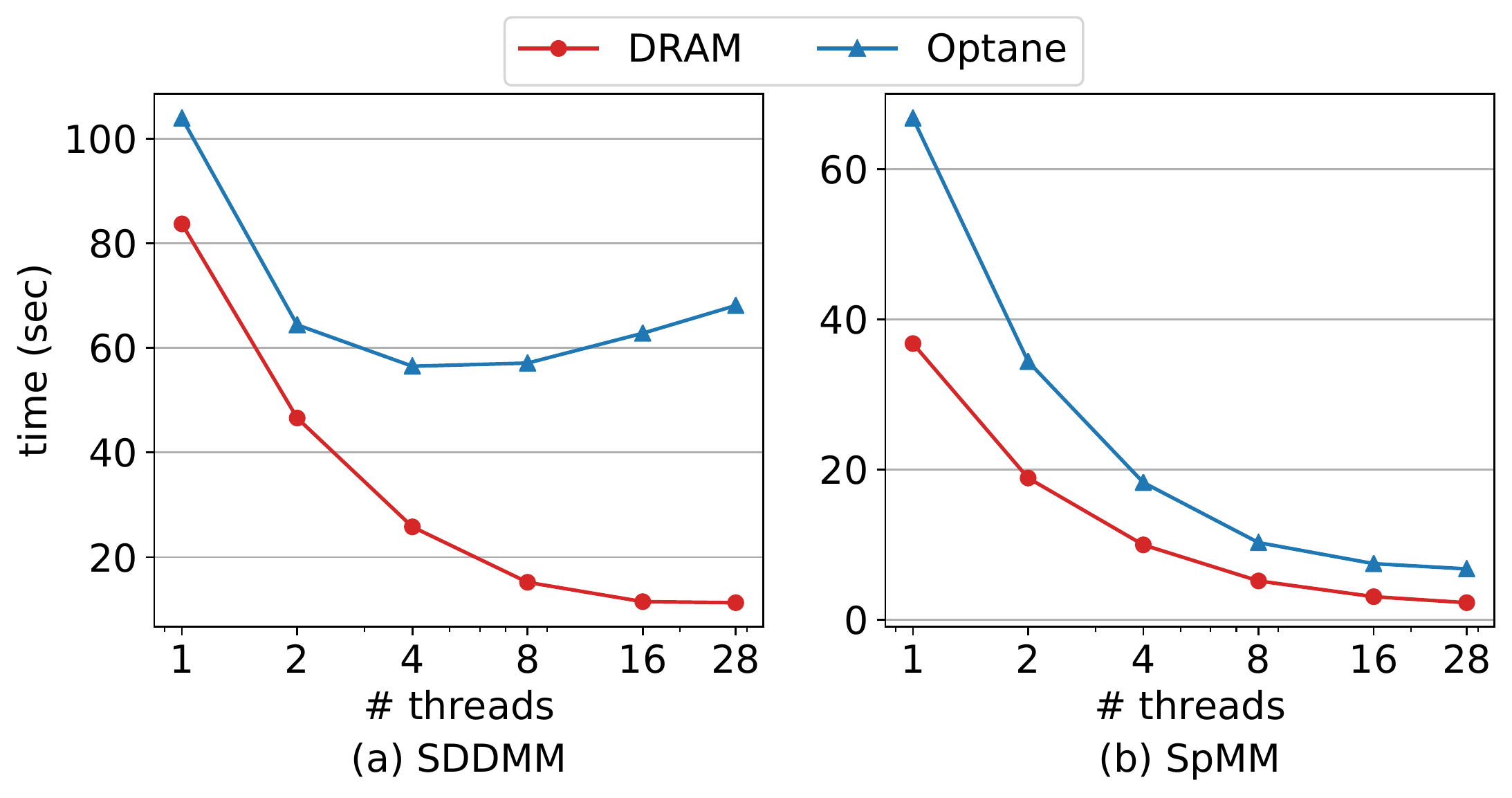}
\vspace{-2em}
\caption{Impact of the number of threads --- \textnormal{The dataset is \texttt{m-x25}; the embedding length is 128.}}
\label{fig:different-thread-numbers-kernel-perf}
\vspace{-1em}
\end{figure}

Figure \ref{fig:different-thread-numbers-kernel-perf} shows the impact of the number of threads on the performance of SDDMM and SpMM, tested on the \texttt{m-x25} dataset.
For SDDMM, the optimal performance is achieved with 4 threads on Optane and 28 threads on DRAM.
This result is consistent with the observation we made in Figure \ref{fig:bandwidth-1} that the nt write bandwidth of Optane saturates at 4--8 threads.
For SpMM, the optimal performance is achieved with 28 threads for both Optane and DRAM, which is consistent with the observation that the read bandwidth of both Optane and DRAM keeps increasing with the number of threads up to 28.
Results on \texttt{g-x256} and \texttt{a-x100} show the same trend.
Ideally, we should set the number of threads to 4 for SDDMM and 28 for SpMM.
However, the DGL framework does not support tuning the number of threads per compute kernel.
In remaining experiments, we still use 28 threads (per socket) for both kernels, which is optimal for SpMM but leads to around 20\% performance drop for SDDMM on Optane compared with using 4 threads.

\begin{figure}[t]
\centering
\includegraphics[width=0.95\linewidth]{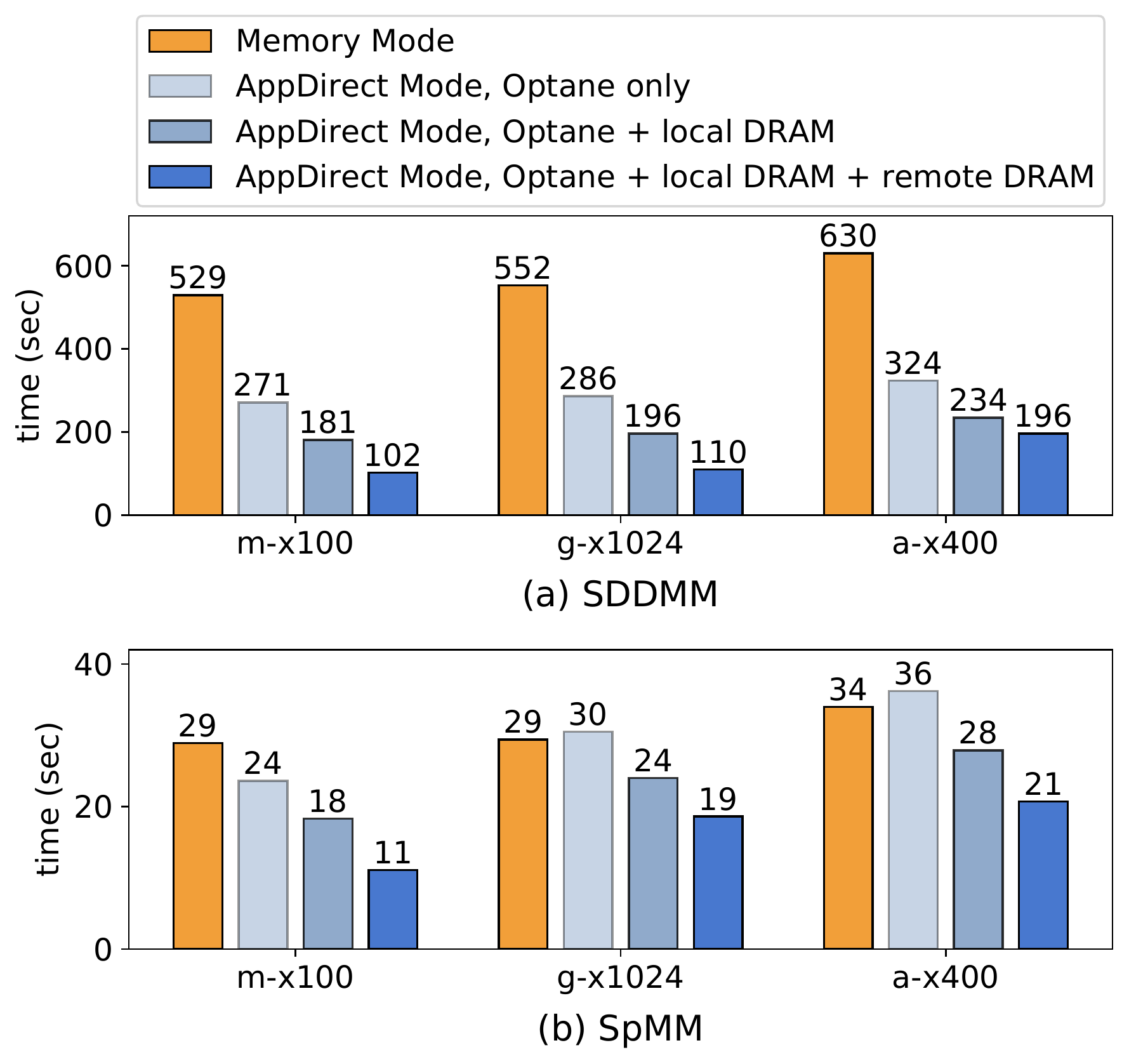}
\vspace{-1em}
\caption{Impact of Optane configurations --- \textnormal{The embedding length is 128; experiments run on a single socket using 28 threads.}}
\label{fig:different-optane-configs-kernel-perf}
\vspace{-1em}
\end{figure}

Figure \ref{fig:different-optane-configs-kernel-perf} shows the impact of different Optane configurations, tested on the three synthetic billion-edge graphs, i.e., \texttt{m-x100}, \texttt{g-x1024}, and \texttt{a-x400}.
Data exceeds the DRAM capacity but fits into Optane on a single socket.
The main results are:
(1) AppDirect Mode performs better than Memory Mode.
Specifically, for SpMM, using Optane$+$DRAM in AppDirect Mode is 1.2--1.6$\times$ faster than Memory Mode; for SDDMM, even using Optane alone is 2$\times$ faster than Memory Mode.
This result indicates that managing DRAM as a cache by the hardware at a granularity of 64 bytes is not suitable for GNNRecSys workloads whose memory access size is hundreds of bytes.
Overhead associated with cache metadata management \cite{hildebrand2021case} could also have contributed to the inferior performance of Memory Mode.
(2) Compared with using Optane alone, using Optane$+$DRAM (in AppDirect Mode) improves the performance by 1.5$\times$ for SDDMM and 1.3$\times$ for SpMM.
(3) Using the DRAM located on the remote socket further improves the performance by 1.2--1.8$\times$ for SDDMM and 1.3--1.6$\times$ for SpMM.
This result is consistent with the observation we made in Figure \ref{fig:bandwidth-3} that the remote nt write bandwidth of DRAM is higher than the local nt write bandwidth of Optane.
This result suggests that we should allocate memory on all the sockets to utilize more
DRAMs even when data fits into Optane on a single socket.
In the remaining experiments, we use AppDirect Mode and first allocate memory on DRAMs.

\begin{figure}[t]
\centering
\includegraphics[width=0.95\linewidth]{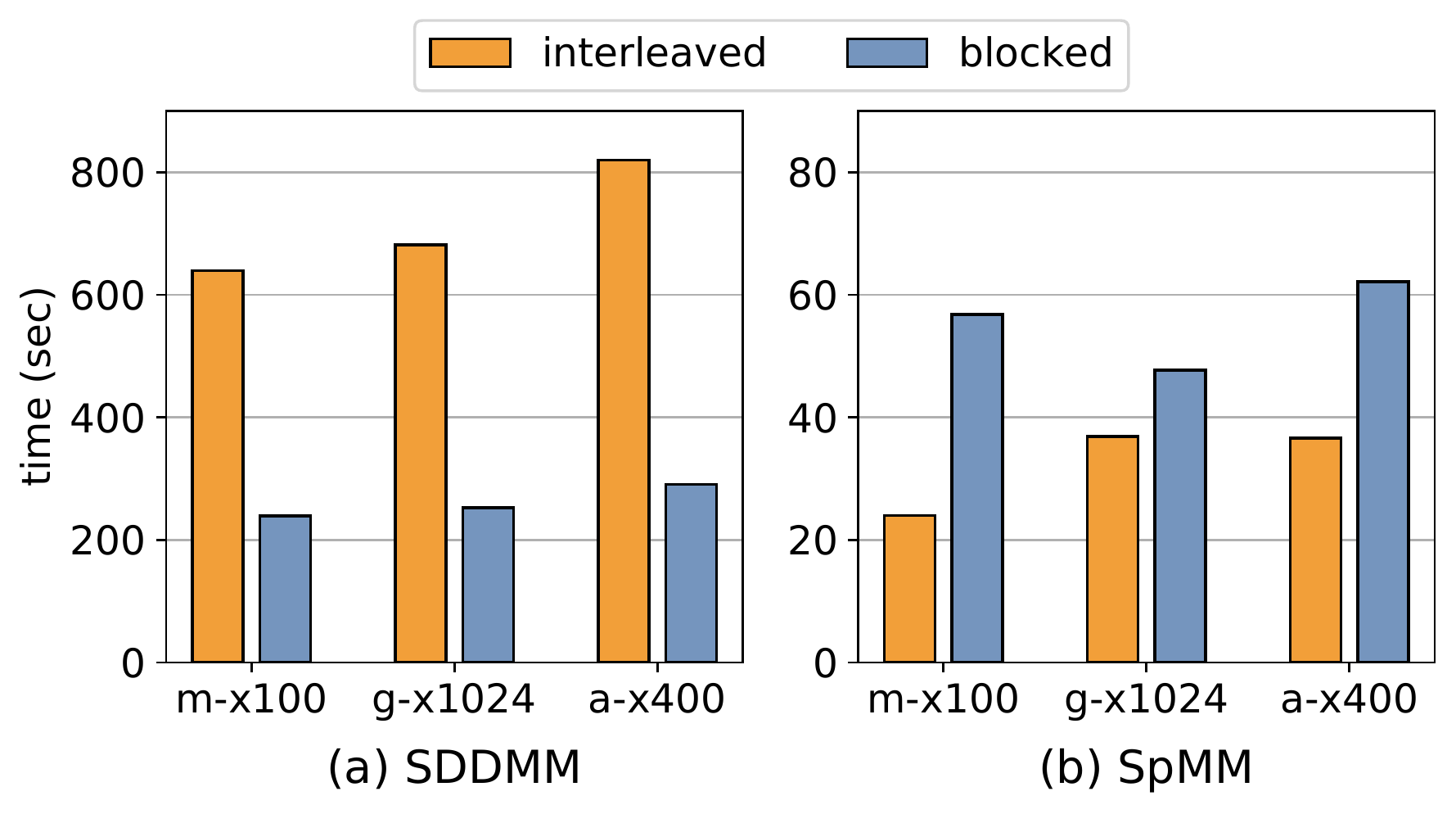}
\vspace{-1em}
\caption{Impact of NUMA data placement policies --- \textnormal{The embedding length is 256; experiments run on two sockets (56 threads).}}
\label{fig:different-numa-data-placement-policies-kernel-perf}
\vspace{-1em}
\end{figure}

Figure \ref{fig:different-numa-data-placement-policies-kernel-perf} shows the impact of NUMA data placement policies.
We consider two policies: (1) interleaving the pages across socket in a round-robin
fashion and (2) blocking the pages and distributing the blocks among sockets.
We find that for SDDMM, blocked data placement works better; for SpMM, interleaved data placement works better.
SDDMM and SpMM prefer different NUMA data placement policies because of their different computation patterns.
For SDDMM, each thread writes to a contiguous range of memory locations; therefore, blocked data placement can guarantee that each thread only needs to access its local socket.
For SpMM, each thread gathers values from incontiguous memory locations and aggregates the values; although interleaved data placement can not fully eliminate cross-NUMA accesses, it incurs fewer than blocked data placement.
For end-to-end experiments, we use blocked NUMA data placement for optimal overall performance.

The above results provide the following guidelines: (1) using nt write for SDDMM and normal write for SpMM; (2) preferring AppDirect Mode over Memory Mode; (3) preferring blocked NUMA data placement over interleaved.
We adopt the three guidelines when doing end-to-end experiments.
Tuning the number of threads for SDDMM and SpMM separately can also be beneficial, but we did not explore this optimization as it is not supported by the DGL framework.

\section{End-To-End Analysis and Optimization}
\label{sec:end2end-benchmarking}

\subsection{Model Accuracy}

In this subsection, we first present techniques for large-batch training.
Then we study the model accuracy of multiple variants of NGCF and LightGCN that have different embedding length and number of layers.
We also evaluate the impact of sampling.

The training setup is as follows.
We split the edges of a graph into 90\% for training and 10\% for testing.
We train the model for 100 epochs.
The evaluation metric is recall@20, i.e., recall calculated based on the top-20 recommended items.
The loss function is Bayesian personalized ranking (BPR) \cite{bpr}, which assumes that observed interactions should be assigned higher prediction values than unobserved ones.
To calculate the BPR loss requires a tuple of a user, one of the user's interacted items, and one un-interacted item.
The original implementation of NGCF and LightGCN computes 1K tuples at each step (hence a batch size of 1K) with a learning rate of 0.0001.

\begin{figure}[t]
\centering
\includegraphics[width=0.75\linewidth]{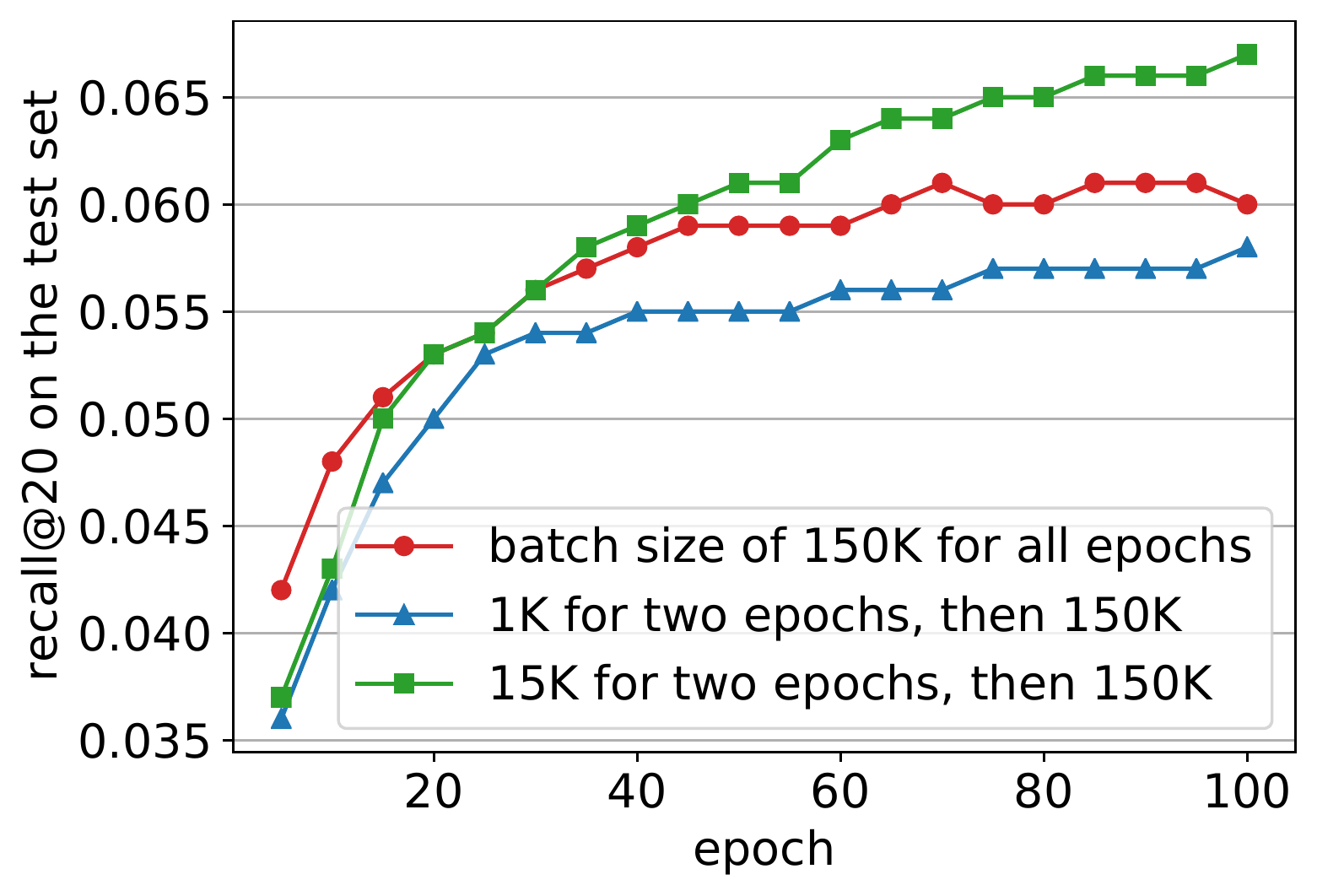}
\vspace{-1em}
\caption{Training curve of LightGCN-2L-128E.}
\label{fig:train-log}
\vspace{-1em}
\end{figure}

For faster training, especially on large graphs, we need to increase the batch size and scale the learning rate accordingly.
We tried both square root scaling \cite{krizhevsky2014one} and linear scaling \cite{imagenet-1-hour}, and found the latter works better.
Furthermore, inspired by the observation made by McCandlish et al. \cite{mccandlish2018empirical} that a large batch size makes training less stable during the first few epochs, we adopt a warm-up training strategy.
Figure \ref{fig:train-log} shows that using a batch size of 15K for the first two epochs and 150K for the remaining epochs achieves a higher recall@20 than using a batch size of 150K throughout all the epochs.
Using a too small batch size for the warm-up stage (1K in this case), however, hurts the model accuracy.
Our empirical study suggests setting the warm-up batch size to 1/10th of the large batch size.
With the aforementioned techniques, we managed to increase the batch size to 150K while achieving the same or even better recall@20 as using the original 1K batch size.

\begin{table}[t]
\caption{Recall@20 of different model variants --- \textnormal{The dataset is \texttt{amazon-book}; a higher recall is better.}}
\label{tab:recall}
\vspace{-1em}
\centering
\begin{subtable}[h]{1.0\linewidth}
\centering
\caption{NGCF}
\vspace{-0.5em}
\begin{adjustbox}{width=0.9\linewidth}
\begin{tabular}{c|ccc}
\toprule
& 1 layer & 2 layers & 3 layers \\
\midrule
\texttt{embedding length = 128}  & 0.065  & 0.072  & 0.077  \\
\texttt{embedding length = 256}  & 0.069  & 0.076  & 0.078  \\
\bottomrule
\end{tabular}
\end{adjustbox}
\end{subtable}
\begin{subtable}[h]{1.0\linewidth}
\centering
\vspace{0.5em}
\caption{LightGCN}
\vspace{-0.5em}
\begin{adjustbox}{width=0.9\linewidth}
\begin{tabular}{c|ccc}
\toprule
& 1 layer & 2 layers & 3 layers \\
\midrule
\texttt{embedding length = 128}  & 0.061  & 0.067  & 0.068  \\
\texttt{embedding length = 256}  & 0.067  & 0.072  & 0.074  \\
\bottomrule
\end{tabular}
\end{adjustbox}
\end{subtable}
\end{table}

\begin{table}[t]
\caption{Recall@20 degradation due to sampling --- \textnormal{The dataset is \texttt{amazon-book}; the model is NGCF-3L-256E.}}
\label{tab:recall-drop-sampling}
\vspace{-1em}
\centering
\begin{adjustbox}{width=0.8\linewidth}
\begin{tabular}{c|cccc}
\toprule
sampling factor & 10     & 20     & 50     & 100 \\
\midrule
degradation     & -0.006 & -0.004 & -0.002 & -0.001 \\
\bottomrule
\end{tabular}
\end{adjustbox}
\end{table}

\begin{figure}[t]
\centering
\includegraphics[width=0.85\linewidth]{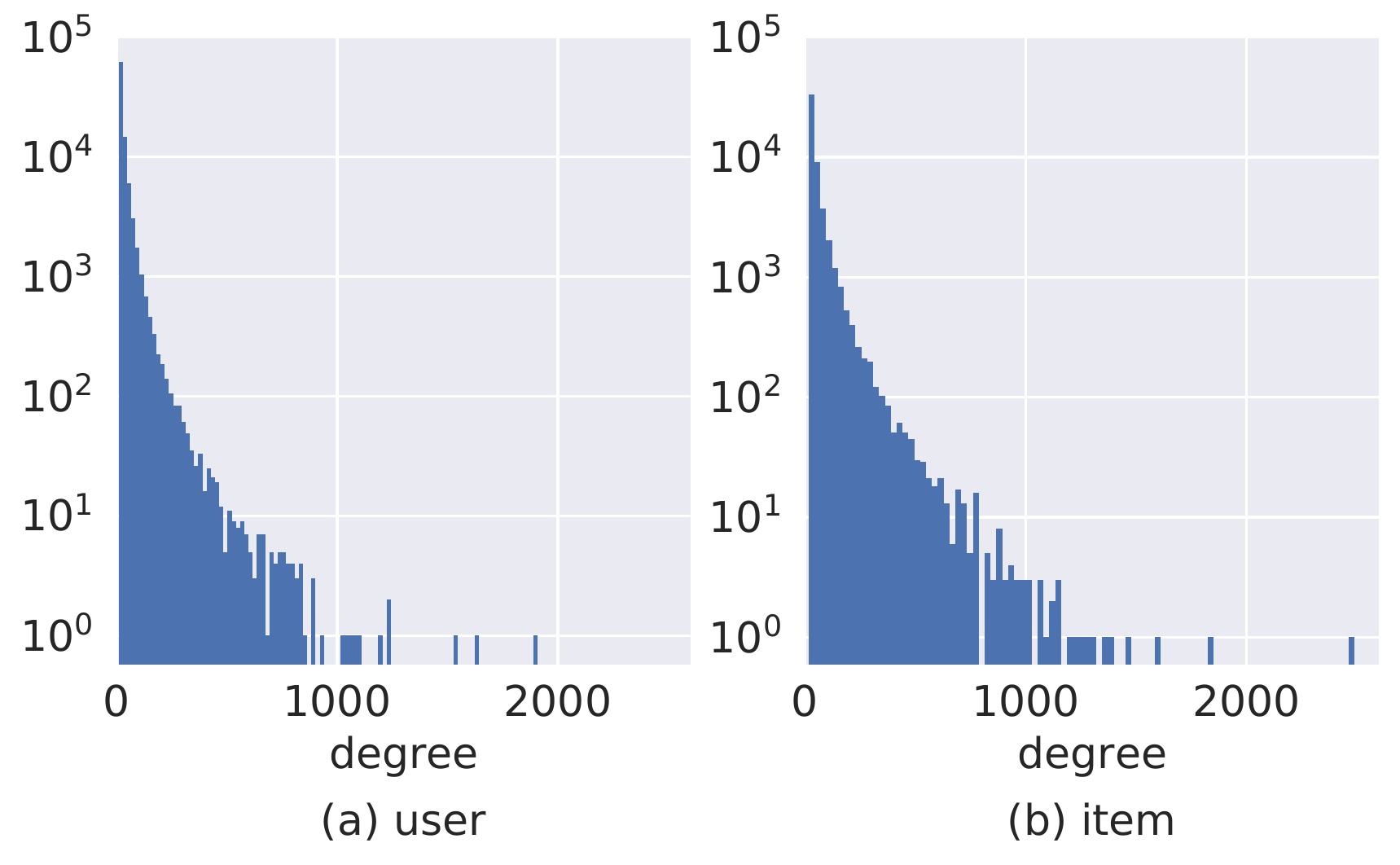}
\vspace{-1em}
\caption{Degree distribution of \texttt{amazon-book}.}
\label{fig:degree-distribution}
\vspace{-1em}
\end{figure}

Table \ref{tab:recall} reports recall@20 of multiple NGCF and LightGCN variants on the \texttt{amazon-book} dataset.
With the same embedding length and number of layers, NGCF outperforms LightGCN.
For both NGCF and LightGCN, increasing either the embedding length or the number of layers leads to improved recall@20.
We see the same trend on \texttt{movielens-10m} and \texttt{gowalla} as well, confirming that a larger GNN model is desired for building higher-quality recommender systems.

Table \ref{tab:recall-drop-sampling} reports recall@20 degradation due to sampling for NGCF-3L-256E.
When the sampling factor is 10 (i.e., sampling 10 neighbors for each vertex), the recall@20 of NGCF-3L-256E degrades from 0.078 to 0.072, which is even worse than the recall@20 of NGCF-2L-256E without sampling.
When the sampling factor is 100, which is larger than the average degree of \texttt{amazon-book}\footnote{When the number of neighbors for a vertex is smaller than the sampling factor, all the neighbors are selected, hence effectively no sampling for the vertex.}, still, a degradation of 0.001 is incurred.
The reason is that \texttt{amazon-book}, as well as other user-item interaction graphs, has a power-law degree distribution where a small portion of vertices have a degree significantly larger than the average, as shown in Figure \ref{fig:degree-distribution}.
Those high-degree vertices (either popular items or active users) tend to be more important for the task of recommendation, but they also suffer more severe information loss from sampling.

\subsection{Comparison With DistDGL}


\begin{table}[t]
\caption{Maximum aggregate batch size allowed by the memory capacity when using DistDGL --- \textnormal{The sampling factor is 100; ``/'' means out-of-memory even at a batch size of one per machine.}}
\label{tab:batch-size-distdgl}
\vspace{-1em}
\centering
\begin{adjustbox}{width=1\linewidth}
\begin{tabular}{c|cccccc}
\toprule
& 1L-128E & 1L-256E & 2L-128E & 2L-256E & 3L-128E & 3L-256E  \\
\midrule
w/o sampling & 24K  & 12K  & 384 & 192 & /   & /   \\
w/ sampling  & 150K & 150K & 12K & 6K  & 256 & 128 \\
\bottomrule
\end{tabular}
\end{adjustbox}
\vspace{-1.5em}
\end{table}

\begin{table*}[t]
\caption{Speedup of Optane-based single-machine training over DistDGL --- \textnormal{Time (unit: sec) is for processing 150K edges.}}
\label{tab:time-one-epoch}
\vspace{-1em}
\centering
\begin{adjustbox}{width=0.75\linewidth}
\begin{tabular}{c|c|c|c|c|c|c}
\toprule
\multicolumn{2}{c|}{} & DistDGL w/o & DistDGL w/ & \multirow{2}{*}{Optane} & {speedup over} & {speedup over} \\
\multicolumn{2}{c|}{} & sampling & sampling & & w/o sampling & w/ sampling \\
\midrule
\multicolumn{1}{c|}{\multirow{3}{*}{NGCF-1L-128E}}
& \texttt{m-x25}    	& 139 & 7	& 100  & 1.4$\times$  & 0.07$\times$ \\
& \texttt{g-x256}   	& 41  & 13	& 103  & 0.4$\times$  &	0.13$\times$ \\
& \texttt{a-x100} 	  	& 45  & 14	& 155  & 0.3$\times$  &	0.09$\times$ \\
\midrule
\multicolumn{1}{c|}{\multirow{3}{*}{NGCF-2L-128E}}
& \texttt{m-x25}    	& 66651 & 92  & 187 & 355.7$\times$	& 0.5$\times$ \\
& \texttt{g-x256}   	& 20434 & 234 & 228 & 89.8$\times$	& 1.0$\times$ \\
& \texttt{a-x100} 	  	& 18626	& 272 & 308 & 60.5$\times$	& 0.9$\times$ \\
\midrule
\multicolumn{1}{c|}{\multirow{3}{*}{NGCF-3L-128E}}
& \texttt{m-x25}    	& / & 9219	  & 287 & /	& 32.1$\times$ \\
& \texttt{g-x256}   	& / & 27209   & 335 & /	& 81.2$\times$ \\
& \texttt{a-x100} 	  	& / & 36054	  & 425 & /	& 84.9$\times$ \\
\midrule

\multicolumn{1}{c|}{\multirow{3}{*}{LightGCN-1L-128E}}
& \texttt{m-x25}    	& 72 & 5	& 60  & 1.2$\times$  & 0.09$\times$ \\
& \texttt{g-x256}   	& 24 & 12	& 70  & 0.3$\times$  & 0.16$\times$ \\
& \texttt{a-x100} 	  	& 29 & 13	& 124 & 0.2$\times$  & 0.11$\times$ \\
\midrule
\multicolumn{1}{c|}{\multirow{3}{*}{LightGCN-2L-128E}}
& \texttt{m-x25}    	& 36847 & 60	& 106  & 347.1$\times$  & 0.6$\times$ \\
& \texttt{g-x256}   	& 11331 & 119	& 140  & 81.1$\times$   & 0.9$\times$ \\
& \texttt{a-x100} 	  	& 10649 & 137	& 251  & 42.5$\times$   & 0.6$\times$ \\
\midrule
\multicolumn{1}{c|}{\multirow{3}{*}{LightGCN-3L-128E}}
& \texttt{m-x25}    	& / & 5440	& 170  & / & 31.9$\times$ \\
& \texttt{g-x256}   	& / & 15264 & 197  & / & 77.5$\times$ \\
& \texttt{a-x100} 	  	& / & 21233 & 357  & / & 59.4$\times$ \\
\bottomrule
\end{tabular}
\end{adjustbox}
\end{table*}

In this subsection, we compare Optane-based single-machine training with distributed training using DistDGL, in terms of hardware cost (\$) and performance.
We perform experiments on \texttt{m-x25}, \texttt{g-x256}, and \texttt{a-x100} datasets.

We run DistDGL on an in-house cluster of six servers.
Each server is a two-socket 32-core 2.8 GHz Intel Xeon Gold 6242 machine with 384 GB DDR4 memory.
The servers are in the same rack and connect to 10 Gigabit Ethernet.
The aggregate memory capacity of the cluster is 2304 GB, which is 20\% larger than the memory capacity of the Optane machine (1536 GB of Optane plus 384 GB of DRAM).
For the cluster, the total price of the DRAMs is \$27K; for the Optane machine, the price of Optane$+$DRAM is \$11K \cite{optane-price}.

Table \ref{tab:batch-size-distdgl} reports the maximum aggregate batch size that is allowed by the memory capacity when using DistDGL.
An aggregate batch size of six means a batch size of one per machine.
The batch size quickly decreases as the number of layers increases.
Specifically, without sampling, NGCF(LightGCN)-1L-128E allows a batch size of 24K, while NGCF(LightGCN)-2L-128E only allows 384; further increasing the number of layers to three would run out of memory even at a batch size of one per machine.
Sampling increases the batch size and makes three-layer NGCF and LightGCN models possible to train.
For NGCF(LightGCN)-1L-128E and NGCF(LightGCN)-1L-256E, we set the batch size to 150K not because of the memory capacity limit, but to ensure training convergence.
All Optane experiments use a batch size of 150K.

\begin{figure}[t]
\centering
\includegraphics[width=0.95\linewidth]{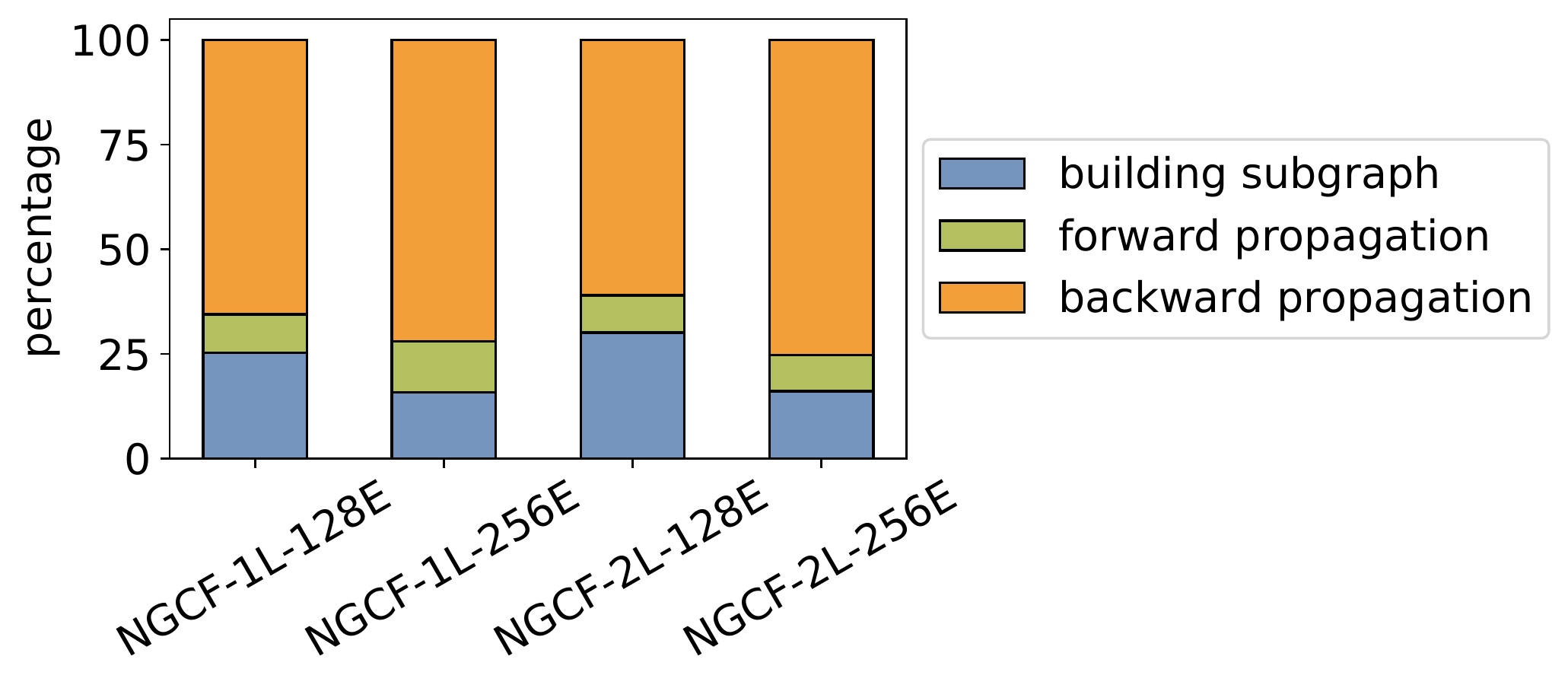}
\vspace{-1em}
\caption{Execution time breakdown of DistDGL --- \textnormal{The dataset is \texttt{m-x25}; w/o sampling; execution time of each part is averaged over six machines.}}
\label{fig:distdgl-time-breakdown}
\vspace{-1.5em}
\end{figure}

Table \ref{tab:time-one-epoch} reports the performance comparison between Optane-based single-machine training and DistDGL, measured by the execution time for processing 150K edges.
The key observation is that Optane-based single-machine training achieves significant speedup over DistDGL on deep NGCF and LightGCN models (with two or three layers), because the execution time of Optane-based single-machine training increases roughly linearly with the number of layers while that of DistDGL exponentially.
For example, on \texttt{m-x25}, using Optane takes 100 seconds for NGCF-1L-128E and 287 seconds for NGCF-3L-128E, while DistDGL with sampling takes 7 seconds for NGCF-1L-128E and 9219 seconds for NGCF-3L-128E.
On NGCF-2L-128E and LightGCN-2L-128E, using Optane brings 43--356$\times$ speedup over DistDGL without sampling.
On NGCF-3L-128E and LightGCN-3L-128E, even with sampling, using Optane still brings up to 85$\times$ speedup.
The results demonstrate the advantages of Optane-based single-machine training for handling deep GNN models.

To further examine the inefficiencies of DistDGL, we plot its execution time breakdown in Figure \ref{fig:distdgl-time-breakdown}.
Building subgraph takes 16--32\% of the total time, even higher than the forward propagation, which takes only about 10\%.
The backward propagation takes more than 60\% of the total time, which is 6$\times$ higher than the forward propagation.
In comparison, for Optane-based single-machine training, the execution time of backward propagation is only 2--3$\times$ higher than forward propagation.
The slowness of the backward propagation in DistDGL is due to two factors: (1) exchanging gradients across machines incurs communication overhead; (2) since DistDGL performs synchronous training, the execution time is determined by the slowest machine.


\section{Discussion}
\label{sec:discussion}

\subsection{Optimization Opportunities}

\emph{Cache Optimization for SDDMM and SpMM.}
There have been attempts to improve cache utilization of SDDMM and SpMM compute kernels through locality-enhancing scheduling.
For example, FeatGraph \cite{featgraph} combines graph partitioning with embedding tiling to strike a balance between efficiency of accessing the graph structure and efficiency of accessing embeddings.
Hong et al. \cite{ASpT} propose an adaptive tiling strategy that takes into account the degree of each vertex.
Various graph reordering methods have also been proposed \cite{rabbit, balaji2018graph, mukkara2018exploiting}.
All these techniques, however, are evaluated on DRAM-only machines, and it is unclear how effective they are on Optane.
Specifically, these techniques are unaware of the asymmetric read and write bandwidth of Optane.
We envision that to achieve optimal performance on Optane, a partitioning or reordering technique should prioritize high cache utilization for write, instead of treating read and write equally.


\emph{Hybrid Memory Management Tailored to GNNRecSys.}
Although this work focused on using Optane in Memory Mode and in AppDirect Mode through existing NUMA utilities, we envision there is room for performance improvement by tailoring the hybrid memory management policy to GNNRecSys.
We observed that turning off NUMA page migration achieves performance comparable to that with NUMA page migration on, or even 5\% higher in certain cases likely due to less time in kernel space.
This observation indicates that managing the hybrid Optane$+$DRAM memory system through existing NUMA utilities is far from optimal for GNNRecSys.
AutoTM \cite{autotm} analyzed inefficiencies of the NUMA approach in training convolutional neural networks (CNNs) and proposed to optimize the location and movement of tensors between DRAM and Optane based on an integer linear programming (ILP) formulation.
AutoTM requires each tensor to fit into DRAM, which is not guaranteed for GNNRecSys when the graph is large.
It is a promising avenue for future research to extend AutoTM to support hybrid memory management at a sub-tensor granularity.




\subsection{Beyond Optane}

To our knowledge, Optane is the only commercially-available persistent memory when we conduct this research.
In this work, we use the first generation of Optane (100 Series) to perform the evaluation.
The second generation (200 Series) provides 32\% higher memory bandwidth; we expect higher performance of GNNRecSys workloads on it.

Although Intel recently announced the discontinuation of future Optane products \cite{intel-kills-optane}, upcoming persistent memory devices from Kioxia and Everspin \cite{Kioxia-and-Everspin} may become suitable alternatives.
On these emerging devices, the key finding of this work should remain valid, that is, single-machine full-graph GNNRecSys training enabled by a large memory capacity outperforms distributed subgraph training.
The techniques for large-batch training also remain useful.
The observations we made in Section \ref{sec:kernel-level-benchmarking} do not necessarily hold, since they depend on the bandwidth characteristics of the specific persistent memory device and how it interacts with DRAMs in the system.
Nevertheless, we can follow the same methodology of this work to analyze and optimize GNNRecSys workloads on new types of persistent memory devices; we will open-source the code to facilitate such process.

Besides persistent memory, Compute Express Link (CXL) \cite{cxl} provides another solution for expanding the memory capacity.
CXL is an open, industry-backed interconnect standard based on PCIe.
With PCIe 5.0, CXL bandwidth is comparable to the cross-socket interconnect on a dual-socket machine.
Since there are no commercially-available CXL memory devices yet, researchers have been using remote NUMA nodes to emulate them \cite{pond, tpp}.
We can view the use of Optane in AppDirect Mode through NUMA utilities as an emulation for CXL memory devices.
We expect that CXL memory devices can bring similar advantages to GNNRecSys training as Optane.

\section{Related Work}
\label{sec:related}


\emph{GNN Characterization and Acceleration.}
Prior efforts for characterizing and accelerating GNN workloads mainly focus on GNN models that are used in vertex classification tasks \cite{hygcn, garg2021understanding, awb-gcn, featgraph, dorylus, roc}, most notably the GCN model \cite{gcn}.
Compared with GCN, NGCF and LightGCN consume much more memory for two reasons.
First, GCN has a simpler message generation function --- just multiplying the embedding vector of the source vertex with a normalization value, which is a constant scalar.
As a result, for GCN, message generation and aggregation can be fused into a single SpMM kernel.
In contrast, NGCF and LightGCN require both SpMM and SDDMM.
Second, GCN typically has two layers, whereas NGCF and LightGCN have more layers (the default setting in their papers \cite{ngcf, lightgcn} is three).
Hence, NGCF and LightGCN, as well as other GNN models for recommender systems in general, have a more urgent demand for a high memory capacity than GCN.


\emph{GNN Dataflow Optimization.}
In this work, we manually optimized the dataflow of NGCF and LightGCN.
Recently, Graphiler \cite{graphiler} is proposed to optimize GNN dataflow at the compiler level.
Graphiler iteratively traverses the dataflow graph to match subgraphs with predefined patterns and replace them with optimized ones.
Graphiler currently has limited coverage on possible dataflow optimizations; for example, it does not support the optimization of reusing SDDMM results that we applied to NGCF and LightGCN.
It is worth further research towards automatic and intelligent GNN dataflow optimization.




\emph{Deep Learning for Recommender Systems.}
Prior to GNNs, neural networks such as multi-layer perceptrons (MLPs) have been applied to recommender systems to replace the hard-coded prediction function (typically dot product).
Efforts along this direction include Wide \& Deep \cite{wide-and-deep}, NCF \cite{ncf}, DLRM \cite{dlrm}, to name a few.
GNNRecSys is complementary to these models, that is, the embeddings that a GNN model generates through message passing can be fed into these models as (part of) the input.
Like GNNRecSys workloads, these models also demand a large amount of memory; existing solutions \cite{zhao2019aibox, mudigere2021software} leverage SSD$+$DRAM.
It remains to be explored what a role Optane or CXL can play there.


\vspace{-0.5em}
\section{Conclusion}
\label{sec:conclusion}
\vspace{-0.5em}


This work analyzes and optimizes GNNRecSys training on persistent memory.
Our experiments show that the large capacity of persistent memory makes it a good fit for GNNRecSys.
With the tuned batch size and optimal system configuration, single-machine GNNRecSys training on persistent memory outperforms distributed training by a large margin, especially when handling deep GNN models.
It is our hope that this paper provides practical guidelines for running GNNRecSys on Optane as well as inspiring future research towards more efficient GNNRecSys on emerging memory devices.



\clearpage

\bibliographystyle{plain}
\bibliography{main}

\end{document}